\documentclass[aps,prd,twocolumn,nofootinbib,superscriptaddress,preprintnumbers,floatfix]{revtex4-1}
\usepackage{amsfonts,amsmath,amssymb,amsthm,braket,nicefrac}
\usepackage{booktabs,array}
\usepackage[utf8]{inputenc}
\usepackage{float,graphicx,hhline}
\usepackage{algorithm,algpseudocode}
\usepackage{appendix}
\usepackage[dvipsnames]{xcolor}
\usepackage{mathtools}
\usepackage{multirow}
\usepackage{tikz-cd}
\usepackage{slashed}
\usepackage{soul}
\usepackage{pifont}
\usepackage[caption=false]{subfig}
\usepackage[export]{adjustbox}
\usepackage[normalem]{ulem}
\usepackage{hyperref}
\usepackage{cleveref}
\usepackage{listings}

\AtBeginDocument{
\heavyrulewidth=.08em
\lightrulewidth=.05em
\cmidrulewidth=.03em
\belowrulesep=.65ex
\belowbottomsep=0pt
\aboverulesep=.4ex
\abovetopsep=0pt
\cmidrulesep=\doublerulesep
\cmidrulekern=.5em
\defaultaddspace=.5em
}

\usepackage{hyperref}%
\hypersetup{
    pdftitle={many pions},
    colorlinks=true,     %
    linkcolor=blue,      %
    citecolor=blue,      %
    filecolor=blue,      %
    urlcolor=blue        %
}

\definecolor{codegreen}{rgb}{0,0.6,0}
\definecolor{codegray}{rgb}{0.5,0.5,0.5}
\definecolor{codepurple}{rgb}{0.58,0,0.82}
\definecolor{backcolour}{rgb}{0.95,0.95,0.92}

\lstdefinestyle{mystyle}{
    backgroundcolor=\color{backcolour},   
    commentstyle=\color{codegreen},
    keywordstyle=\color{magenta},
    numberstyle=\tiny\color{codegray},
    stringstyle=\color{codepurple},
    basicstyle=\ttfamily\footnotesize,
    breakatwhitespace=false,         
    breaklines=true,                 
    captionpos=b,                    
    keepspaces=true,                 
    numbers=none,                    
    numbersep=5pt,                  
    showspaces=false,                
    showstringspaces=false,
    showtabs=false,                  
    tabsize=2
}

\DeclareMathAlphabet{\mathbbold}{U}{bbold}{m}{n}

\DeclareMathOperator{\Tr}{Tr}

\begin{document}

\title{Lattice quantum chromodynamics at large isospin density: 6144 pions in a box}

\newcommand{\getMITAffiliation}{\affiliation{Center for Theoretical Physics, Massachusetts Institute of Technology, Cambridge, MA 02139, USA}}
\newcommand{\getIAIFIAffiliation}{\affiliation{The NSF AI Institute for Artificial Intelligence and Fundamental Interactions}}

\author{Ryan Abbott}
\email[Corresponding author: ]{rabbott@mit.edu}
\getMITAffiliation
\getIAIFIAffiliation
\author{William Detmold}
\getMITAffiliation
\getIAIFIAffiliation
\author{Fernando Romero-L\'opez}
\getMITAffiliation
\getIAIFIAffiliation

\author{Zohreh~Davoudi}
\address{Department of Physics and Maryland Center for Fundamental Physics, University of Maryland, College Park, MD 20742, USA}
\address{Joint Center for Quantum Information and Computer Science, NIST/University of Maryland, College Park, MD 20742, USA}

\author{Marc~Illa}
\affiliation{InQubator for Quantum Simulation (IQuS), Department of Physics, University of Washington, Seattle, WA 98195, USA}

\author{Assumpta~Parre\~no}
\affiliation{Departament de F\'{\i}sica Qu\`{a}ntica i Astrof\'{\i}sica and Institut de Ci\`{e}ncies del Cosmos,	Universitat de Barcelona, Mart\'{\i} i Franqu\`es 1, E08028, Spain}

\author{Robert J. Perry}
\affiliation{Departament de F\'{\i}sica Qu\`{a}ntica i Astrof\'{\i}sica and Institut de Ci\`{e}ncies del Cosmos,	Universitat de Barcelona, Mart\'{\i} i Franqu\`es 1, E08028, Spain}

\author{Phiala~E.~Shanahan } 
\getMITAffiliation
\getIAIFIAffiliation

\author{Michael~L.~Wagman}
\affiliation{Fermi National Accelerator Laboratory, Batavia, IL 60510, USA}

\collaboration{NPLQCD collaboration}

\preprint{MIT-CTP/5560,UMD-PP-023-03,FERMILAB-PUB-23-382-T}

\begin{abstract}
We present an algorithm to compute correlation functions for systems with the quantum numbers of many identical mesons from lattice quantum chromodynamics (QCD). The algorithm is numerically stable and allows for the computation of $n$-pion correlation functions for $n \in \{ 1, \dots, N\}$ using a single $N \times N$ matrix decomposition, improving on previous algorithms.  We apply the algorithm to calculations of correlation functions with up to 6144 charged pions using two ensembles of gauge field configurations generated with quark masses corresponding to a pion mass $m_\pi =  170$ MeV and spacetime volumes of $(4.4^3\times 8.8)\ {\rm fm}^4$ and $(5.8^3\times 11.6)\ {\rm fm}^4$. We also discuss statistical techniques for the analysis of such systems, in which the correlation functions vary over many orders of magnitude. In particular, we observe that the many-pion correlation functions are well approximated by log-normal distributions, allowing the extraction of the energies of these systems. Using these energies, the large-isospin-density, zero-baryon-density region of the QCD phase diagram is explored. A peak  is observed in the energy density at an isospin  chemical potential $\mu_I\sim 1.5 m_\pi$, signalling the transition into a Bose-Einstein condensed phase. The isentropic speed of sound, $c_s$, in the medium is seen to exceed the ideal-gas (conformal) limit ($c_s^2\leq 1/3$) over a wide range of chemical potential before falling towards the asymptotic expectation at $\mu_I\sim 15 m_\pi$. These, and other thermodynamic observables, indicate that the isospin chemical potential must be large for the system to be well described by an ideal gas or perturbative QCD.
\end{abstract}
\maketitle

\section{Introduction}

Describing strongly-interacting dense matter is a central challenge for nuclear physics. Whilst the strong interactions are governed by quantum chromodynamics (QCD), the many-body nature of these interactions in dense environments such as neutron stars, supernovae, and binary mergers makes the task of predicting the behavior of these systems exceedingly difficult. Numerical studies at non-zero baryon density or chemical potential using lattice QCD (LQCD) are frustrated by a sign problem that prohibits efficient stochastic evaluations of the integrals that define physical observables. Consequently, most studies of the neutron star equation of state (EoS), for example, use effective models or interpolate between limited phenomenological inputs at small values of the chemical potential, and limits from perturbative QCD (pQCD) at asymptotically large chemical potential. In contrast, systems with non-zero isospin chemical potential (in which up and down quarks have opposite values of their chemical potentials), denoted by $\mu_I$, are amenable to LQCD calculations~\cite{Kogut:2002tm,Kogut:2002zg,Kogut:2004zg,Endrodi:2014lja,deForcrand:2007uz,Cea:2012ev,Brandt:2017oyy,Brandt:2018omg,Brandt:2023kev,Brandt:2022hwy,Detmold:2008fn,Detmold:2012pi,Detmold:2012pi2}. Pure isospin-chemical-potential systems are not directly relevant to neutron stars and other known astrophysical objects as they have non-zero values of both baryon- and isospin-chemical potential.\footnote{The possibility of pion stars has also been conjectured \cite{Brandt:2018bwq}.} Nonetheless, these systems can provide an interesting testing ground for effective models and asymptotic pQCD expectations and hence it is interesting to seek first-principles QCD predictions.

LQCD studies at non-zero isospin chemical potential and zero baryon chemical potential have been performed by adding an explicit chemical potential term $\mu_I(\bar u\gamma_0 u - \bar d \gamma_0 d)$ (where $u$ and $d$ are quark fields, $\gamma_0$ is a Dirac matrix and $\mu_I$ quantifies the size of the chemical-potential) to the QCD action~\cite{Kogut:2002tm,Kogut:2002zg,Kogut:2004zg,Endrodi:2014lja,deForcrand:2007uz,Cea:2012ev,Brandt:2017oyy,Brandt:2018omg,Brandt:2023kev,Brandt:2022hwy}. Additionally, a canonical approach to isospin chemical potential is enabled by studies of systems of fixed isospin charge (fixed numbers of charged pions) in a finite box~\cite{Detmold:2008fn,Detmold:2012pi,Detmold:2012pi2}; these systems are characterized by large isospin density, and hence probe the large isospin-chemical-potential, zero baryon-chemical-potential, region of the QCD phase diagram. Two transitions are expected at zero temperature as $\mu_I$ increases: a first-order phase transition at $|\mu_I| = m_\pi$ from a weakly-interacting pion gas to a Bose-Einstein condensate (BEC) phase, and a crossover at larger $\mu_I$ from the BEC phase to a deconfined superconducting Bardeen-Cooper-Schrieffer (BCS) phase~\cite{son_qcd_2001}; see Ref.~\cite{Mannarelli:2019hgn} for a review. Previous work on QCD at non-zero isospin chemical potential using both approaches has found evidence for the BEC transition, but did not definitively probe the BCS phase or the onset of pQCD. 

In this work, we study systems with the quantum numbers of $n$ identical charged pions with zero total three-momentum. Such systems have been investigated in previous work~\cite{Beane:2007qr,Detmold:2008fn,Detmold:2010au,Detmold:2011kw,Detmold:2012pi,Detmold:2014iha,Detmold:2012pi2} 
for  $n\leq 72$~\cite{Detmold:2012pi2}. In order to extend those calculations, we develop an algorithm that allows efficient computation of correlation functions for larger $n$ and numerically investigate $n\leq 6144$ systems. The algorithm is built upon properties of the symmetric group and importantly can be implemented without the extreme numerical-precision requirements of existing approaches.
While the algorithm is specific to the highly symmetric systems that are considered, symmetry-group-based generalizations may be appropriate for efficiently performing the Wick contractions for other multi-hadron systems such as nuclei.

The LQCD two-point correlation functions used to access these $n$-pion systems rapidly decay as the separation of the points increases. This presents numerical challenges in the calculation and analysis of the correlation functions investigated in this work.
The correlation functions not only vary by many orders of magnitude across the lattice extent,
but even at the same site they fluctuate by orders of magnitude between configurations. Consequently, any attainable statistical sample of a many-pion correlation function will be far from the realm of validity of the Central Limit Theorem (CLT), and the statistical estimators used in most LQCD calculations, such as the sample mean and standard deviation, will not be meaningful. The distributions of many-particle correlation functions that are positive-definite on all field configurations in a range of contexts have been found to be approximately log-normal~\cite{Guagnelli:1990jb,MJSPC,endres_noise_2011,DeGrand:2012ik} and that behavior is also found for the many-pion systems studied here. Therefore, to extract physical quantities from the LQCD calculations, we perform an analysis that is based on the empirically-motivated assumption of log-normality. LQCD investigations of nuclei and other many-body systems encounter similar, but not identical, statistical challenges \cite{Beane:2009kya,Beane:2009gs,Beane:2009py,Wagman:2016bam,Wagman:2017gqi,Detmold:2018eqd,Davoudi:2020ngi} and some of the techniques explored here may have more general applicability.

Given these techniques and improvements, this work provides new insights into properties of matter at significantly larger isospin densities than previously studied. In particular, we find that the isentropic speed of sound, $c_s$, exceeds the conformal limit of $c_s^2\leq 1/3$ over a wide range of isospin chemical potential. The results also demonstrate the regime of validity of pQCD in describing isospin-chemical-potential matter is bounded below by $\mu_I\sim 15 m_\pi$. 

The structure of this paper is as follows. In Sec. \ref{sec:algorithm}, a new algorithm for computing many-pion correlation functions that forms the basis of this work is introduced. In Sec. \ref{sec:lattice}, the details of the LQCD calculations that are performed and the basic properties of the resulting correlation functions are presented. Section \ref{sec:analysis} presents an analysis of the statistical properties of these correlation functions and introduces the tools with which their distributions are analysed under the assumption of log-normality to extract physical information about these systems. The physical quantities that are determined from these correlation functions relate to large isospin chemical potential and are discussed in Sec. \ref{sec:isospin}. A brief summary is given in Sec. \ref{sec:summary}. Additional details of numerical tests of the algorithms and data presentations used herein, and investigations of the inclusion of higher cumulants in the analysis, are presented in the Appendices.

\section{Many-pion correlation functions}    
\label{sec:algorithm}

In order to extract physics from LQCD calculations, suitable correlation functions must be constructed and evaluated. In the context of this work, the correlation functions of interest are those that access states with a  large $z$-component of isospin, $I_z$, with vanishing total three-momentum.
 Specifically, we consider
correlation functions of the form
\begin{equation}
\label{eq:correlation-function-defn}
   C_n(t) = \left\langle
    \left(\sum_{x} \pi^-({\bf x}, 0)\right)^n \prod_{i=1}^n 
    \pi^+({\bf y}_i, t)
    \right\rangle,
\end{equation}
where $n=I_z$ labels the minimum number of charged pions required to form the state, and ${\bf y}_i$ are (possibly distinct) spatial lattice sites (the dependence of $C_n(t)$ on these coordinates is suppressed since it does not affect the spectrum of states that propagate over a Euclidean time-separation, $t$). 
Here, $\pi^-({\bf x},t)=\pi^+({\bf x},t)^\dagger = -\overline{d}({\bf x},t) \gamma_5 u({\bf x},t)$, so the sink interpolating field at $t=0$ projects the system to zero total three-momentum by forcing each $\overline{d} u$ pair to zero three-momentum. 

For large $n$, the correlation functions in Eq.~\eqref{eq:correlation-function-defn} involve many quark fields, and, after integration over the fermion degrees of freedom, produce a factorially large set of Wick contractions that must be evaluated and averaged over an ensemble of gluon field configurations. For example, for the largest isospin-charge that we consider in this work, the required number of Wick contractions is $(6144!)^2\sim\mathcal{O}(10^{40000})$. A naive approach to the evaluation of Eq.~\eqref{eq:correlation-function-defn} is therefore impractical for all but small $n$, and more efficient methods are required. As Eq.~\eqref{eq:correlation-function-defn} provides a prototypical (and particularly simple) example of a many-body system, significant effort has been devoted to developing such algorithms. In Refs.~\cite{Beane:2007qr,Detmold:2008fn}, a method based on the expansion of determinants was introduced and used to study $n\leq 12$ pion systems. A more powerful recursive algorithm was introduced in Ref.~\cite{Detmold:2010au} and further developed in Ref.~\cite{Detmold:2012pi2}, allowing $n\leq72$ pion systems to be studied. While these methods were significant steps forward in the study of many-hadron systems in LQCD, they suffer from numerical instabilities when $n$ becomes large and  their scaling with $n$ makes it impractical to study still larger values of $n$.

Here, we introduce a new algorithm based on the representation theory of the symmetric group and formalize a relation introduced in Ref.~\cite{Detmold:2014iha}. In this section, we show the algorithm is numerically stable and more efficient than previous algorithms and consider some generalizations of the approach. 
The improved efficiency and stability make it possible to increase the isospin charge that is studied by multiple orders of magnitude over previous work. 

\subsection{Symmetric polynomial algorithm}\label{sec:sym-poly-algorithm}

As in Refs.~\cite{Beane:2007qr,Detmold:2008fn}, a \emph{zero-momentum pion block} can be defined as
\begin{multline}
\label{eq:pion-block-def}
  \Pi_{(i,\alpha)(j,\beta)}(\mathbf{x}, \mathbf{y};t) \\
  = \sum_{k,\gamma,\mathbf{z}}
  S_{(i,\alpha)(k,\gamma)}(\mathbf{x},0; \mathbf{z},t)
  S^\dag_{(k,\gamma)(j,\beta)}(\mathbf{y},0; \mathbf{z},t)\,,
\end{multline}
where $S(x;y)$ is a quark propagator from $x=({\bf x},t_x)$ to $y=({\bf y},t_y)$. Here, $\{\alpha,\beta,\gamma\}$ and $\{i,j,k\}$ indicate spin and color indices respectively, while $\mathbf{x},\mathbf{y}$, and $\mathbf{z}\in \Lambda_3$ indicate spatial positions selected from a set of lattice sites $\Lambda_3$ which can be the entire spatial lattice geometry or some subset.\footnote{Note that the summed spatial location ${\bf z}$ in Eq.~\eqref{eq:pion-block-def} can in principle range over a different set of spatial sites than the external sites ${\bf x}$ and ${\bf y}$.} This object is a matrix in its $N_s$ spin, $N_c$ color, and $N_\Lambda=\dim(\Lambda_3)$ spatial indices

By combining spin, color, and spatial index labels, the pion block can be recast as a time-dependent $N\times N$ matrix, $\Pi(t)$, where  $N=N_c N_s N_\Lambda$. Since the manipulations below will be independent of the temporal coordinate, the time-dependence of $\Pi(t)$ will be suppressed.
Let $\vec x=\{x_1,\dots, x_N\}$  denote the set of eigenvalues of $\Pi$. 
As we will show in Sec. \ref{sec:proof}, the correlation function in Eq.~\eqref{eq:correlation-function-defn} can be written for $1\leq n\leq N$ as
\begin{equation}
  \label{eq:correlator-formula}
  C_n(t) = n!\, E_n(\vec x),
\end{equation}
where $E_n(\vec x)$ is a homogeneous, degree-$n$, symmetric polynomial over the eigenvalues:
\begin{equation}
\label{eq:Enpoly}
    E_n(\vec x)\equiv E_n(\{x_1, \dots, x_N\}) \equiv \sum_{i_1 < \dots < i_n}^N x_{i_1} \dots x_{i_n}\,,
\end{equation}
where the indices $i_k$ range from $1$ to $N$.
For example, $E_2(\{x_1, x_2, x_3\}) = x_1 x_2 + x_1 x_3 +
x_2 x_3$  and the special cases $E_1(\{x_1,\ldots, x_N\})=\sum_{i=1}^Nx_i = {\rm Tr}(\Pi)$ and $E_N(\{x_1,\ldots, x_N\})=\prod_{i=1}^Nx_i = {\rm Det}(\Pi)$ reduce to known results.

Although Eq.~\eqref{eq:correlator-formula} is
conceptually simple (and was written down in Ref.~\cite{Detmold:2014iha}), directly computing the $\binom{N}{n}$ terms in
the sum in Eq.~\eqref{eq:Enpoly} is computationally intractable for even moderate $N$ and $n$. 
However, $E_n(x_1, \dots x_N)$ can be computed using the following recursive relation\footnote{This recursive relation can be seen directly from Eq.~\eqref{eq:Enpoly} or as a specific case of the more general methods in Ref.~\cite{demmel2006accurate} for computing Schur polynomials.}:
\begin{equation}
\begin{aligned}
  E_k(\{x_1, \dots, x_M\}) =&\ x_M E_{k - 1}(\{x_1, \dots x_{M - 1}\}) \\
  &+ E_k(\{x_1, \dots, x_{M - 1}\}),
  \end{aligned}
\end{equation}
where $E_k({x_1, \dots x_M})=0$ if $M<k$.
By either recursively computing $E_n$ and caching the result, or
building a lookup table, the computational effort needed
to compute the correlation function from the eigenvalues is reduced to $O(N^2)$. In
practice, this computation is effectively of negligible cost since obtaining the
eigendecomposition of $\Pi$ is an $O(N^3)$ operation that dominates
the cost of obtaining the correlation function from a given set of quark propagators.

\subsection{Generalizations}
Although the method as described is particular to systems of positively (or negatively) charged pions, it can be readily generalized to any type of meson correlation function whose contractions do not admit disconnected diagrams by changing the construction of the pion block, Eq.~(\ref{eq:pion-block-def}). For instance, in order to compute a maximal isospin many-kaon correlation function, the pion block $\Pi$ would be replaced by the kaon block defined by
\begin{multline}
    \kappa_{(i,\alpha)(j,\beta)}(\mathbf{x}, \mathbf{y}, t) \\
  = \sum_{k,\gamma,\mathbf{z}}
  S_{u;(i,\alpha)(k,\gamma)}(\mathbf{x},0; \mathbf{z},t)
  S^\dag_{s;(k,\gamma)(j,\beta)}(\mathbf{y},0; \mathbf{z},t),
  \label{eq:kappa}
\end{multline}
where $S_{u}$ refers to the light quark propagator and $S_{s}$ refers to the strange quark propagator. Additionally, the methods in Ref.~\cite{Detmold:2012pi} can be used to evaluate multi-species correlation functions, such as mixed systems of pions and kaons or systems of pions wherein some pions have nonzero momentum. Similarly, the method can also be applied to baryons in  $N_c=2$ QCD~\cite{Detmold:2014qqa,Detmold:2014kba}.

\subsection{Calculation of eigenvalues}\label{sec:calculating-eigenvalues}

Since the pion block $\Pi(t)=S S^\dag$ in Eq.~\eqref{eq:pion-block-def} is explicitly Hermitian and positive-definite, its eigenvalues are equal to its singular values and can be computed using a singular-value decomposition (SVD). Moreover, its eigenvalues are by definition the squares of the singular values of the (spin-colour-spatial) matrix $S$, so the eigenvalues of $\Pi$ may be computed via a singular value decomposition of $S$ directly without explicitly forming $\Pi$. Determining the eigenvalues of $\Pi$ this way is both more efficient and more numerically stable; the condition number of $S$ is approximately the square root of that of $\Pi$, and hence this is the approach that we use throughout this work.

Notably, computing the eigenvalues of $\Pi$ in this manner is quite specific to the case of zero-momentum pions -- in generalizations to blocks of non-zero momentum or for blocks built from non-degenerate quark-anti-quark pairs, the corresponding block object, e.g., $\kappa$ in Eq.~\eqref{eq:kappa}, will not have a positive definite (or even real) spectrum. In these cases, an alternative eigendecomposition method such as the QR algorithm must be used to determine the input to the symmetric-polynomial algorithm. Care must be taken in such scenarios to ensure numerical accuracy as non-Hermitian eigendecompositions are less numerically stable than SVD; it is likely that for non-positive-definite blocks, higher precision arithmetic would be needed in all stages beginning with the formation of the block and possibly even in the linear solve needed to obtain the propagators.

\subsection{Comparison to existing methods}
The LQCD correlation functions that result from applying the symmetric polynomial algorithm are identical to  those calculated using other methods \cite{Beane:2007qr,Detmold:2008fn,Detmold:2010au,Detmold:2012pi}. The only differences are in decreased computational complexity and improved numeric stability. Improvements in computational complexity result from the ability to compute all $N$ possible pion correlation functions from an $N \times N$ pion block with only a single $O(N^3)$ matrix operation to find the eigenvalues; other methods require separate $O(N^3)$ operations for each number of pions, resulting in an overall $O(N^4)$ cost for computing the full set of $C_n$ for $n \in \{ 1, \dots, N\}$.\footnote{Here, we assume that each matrix operation (e.g., matrix multiplication, SVD) takes $O(N^3)$ floating-point operations, which is the case for the implementations used in this work.}

The numerical stability of our method is also significantly improved -- other methods require high-precision arithmetic in order to counteract catastrophic cancellations between terms. Provided that the eigenvalues $\vec{x}$ are positive in Eq.~(\ref{eq:correlator-formula}), there is no possibility for cancellation, and hence high-precision arithmetic is not needed at any point in the calculation. Indeed, we have checked our method against other methods for small systems and found that our double-precision results match those of other methods that required the use of higher-precision floating-point numbers.

\subsection{Proof}
\label{sec:proof}
In order to prove Eq.~(\ref{eq:correlator-formula}), we start from the result of Ref. \cite{Detmold:2008fn} which expresses the correlation function as\footnote{The expression here is generalized from the $N=12$ case shown in Ref. \cite{Detmold:2008fn} and differs in normalization by a factor of $1/(N - n)!$.}
\begin{equation}
    C_n = \frac{1}{(N - n)!}
    \epsilon^{\alpha_1 \dots \alpha_n \xi_1 \dots \xi_{N - n}}
    \epsilon_{\beta_1 \dots \beta_n \xi_1 \dots \xi_{N - n}}
    \Pi_{\alpha_1}^{\beta_1} \dots \Pi_{\alpha_n}^{\beta_n},
\end{equation}
where Greek indices $\alpha_i, \beta_i, \xi_i$ indicate combined spin-color-spatial indices. This can be simplified using the identity
\begin{equation}
    \epsilon^{\alpha_1 \dots \alpha_n \xi_1 \dots \xi_{m}}
    \epsilon_{\beta_1 \dots \beta_n \xi_1 \dots \xi_{m}}
    = m! \sum_{\sigma \in S_n} \epsilon(\sigma)
    \delta^{\alpha_{\sigma(1)}}_{\beta_1} \dots
    \delta^{\alpha_{\sigma(n)}}_{\beta_n},
\end{equation}
where the summation runs over the permutations $\sigma$ of the symmetric group $S_n$ and $\epsilon(\sigma)$ is the sign of the particular permutation. This gives
\begin{equation}
\label{eq:correlator-pi-sigma-sum}
    C_n = \sum_{\sigma \in S_n} \epsilon(\sigma)
    \Pi_{\alpha_1}^{\alpha_{\sigma(1)}}
    \dots \Pi_{\alpha_n}^{\alpha_{\sigma(n)}}.
\end{equation}
Each summand in Eq.~\eqref{eq:correlator-pi-sigma-sum} can be written as a product of traces of powers of the pion block $\Pi$, with the composition of these traces determined by the conjugacy class of $\sigma$. More concretely, if $\lambda(\sigma) = (\lambda_1, \dots, \lambda_q)$ denotes the partition associated to the conjugacy class of $\sigma$ (so $\sigma$ has cycles of length $\lambda_1, \dots, \lambda_{q}$), then Eq.~\eqref{eq:correlator-pi-sigma-sum} may be rewritten as
\begin{equation}
\label{eq:correlator-sum-trace}
    C_n = \sum_{\sigma \in S_n} \epsilon(\sigma)
    \prod_{i = 1}^q \Tr(\Pi^{\lambda_i}).
\end{equation}
The traces can each be written in terms of sums of powers of the elements of the set of eigenvalues $\vec{x}=\{x_1, \dots x_N\}$  of $\Pi$ so
\begin{equation}
  \label{eq:correlator-trace-polynomial}
    C_n = \sum_{\sigma \in S_n} \epsilon(\sigma) P_{\lambda(\sigma)}(\vec x),
\end{equation}
where $P_{\lambda}(\vec x)$ is a \emph{power-sum symmetric polynomial} 
\begin{equation}
  \label{eq:power-sum-poly-def}
    P_\lambda(\vec x) = \prod_{i = 1}^q (x_1^{\lambda_i} + \dots + x_N^{\lambda_i}).
\end{equation}
We can simplify Eq.~\eqref{eq:correlator-trace-polynomial} by changing
from the basis  of symmetric polynomials $P_\lambda(\vec x)$
into the basis of \emph{Schur polynomials}, $S_\lambda(\vec x)$ (not to be confused with the quark propagator), using the
Frobenius Character Formula \cite{alma990011061800106761}\cite[p.~49]{fulton2013representation}
\begin{equation}
    P_\lambda(\vec x) = \sum_{\lambda'} \chi_{\lambda'}(\mathcal{C}_\lambda) S_{\lambda'}(\vec x),
\end{equation}
where $\chi_{\lambda'}(\mathcal{C}_\lambda)$ is the character for the irreducible
representation of the symmetric group associated to the partition $\lambda'$ applied to the conjugacy class $\mathcal{C}_\lambda$ of the partition $\lambda$.

Three additional facts from representation theory are needed to complete the proof; firstly
that $S_{(1,\dots, 1)}(\vec x) = E_n(\vec x)$; secondly, that the character of the $n$-partition $(1,\ldots,1)$ is the sign function, $\chi_{(1,\dots,1)} = \epsilon$; and finally, that the orthogonality relation for characters
\begin{equation}
  \frac{1}{n!} \sum_{\sigma \in S_n}
  \overline{\chi_{\lambda}(\sigma)} \chi_{\lambda'}(\sigma)
  = \delta_{\lambda,\lambda'}
\end{equation}
holds (here the bar indicates complex conjugation) \cite{alma990011061800106761,fulton2013representation}. 

Using these facts, Eq.~\eqref{eq:correlator-formula} follows through a
straightforward computation:
\begin{align}
  C_n
  &= \sum_{\sigma \in S_n} \epsilon(\sigma) P_{\lambda(\sigma)}(\vec x)
	\notag \\
  &= \sum_{\sigma \in S_n}
	\sum_{\lambda'} \epsilon(\sigma) \chi_{\lambda'}(\sigma)
	S_{\lambda'}(\vec x) \notag \\
  &= \sum_{\lambda'} S_{\lambda'}(\vec x) \sum_{\sigma \in S_n}
	\chi_{(1,\ldots,1)}(\sigma) \chi_{\lambda'}(\sigma) \notag \\
  &= n!\, S_{(1, \dots, 1)}(\vec x) \notag \\
  &= n!\, E_n(\vec x)\,.
	\label{eq:correlator-calculation}
\end{align}

\section{Numerical results}

\begin{figure}[t!]
    \centering
    \vspace{-0.1cm}
    \includegraphics[width=\linewidth]{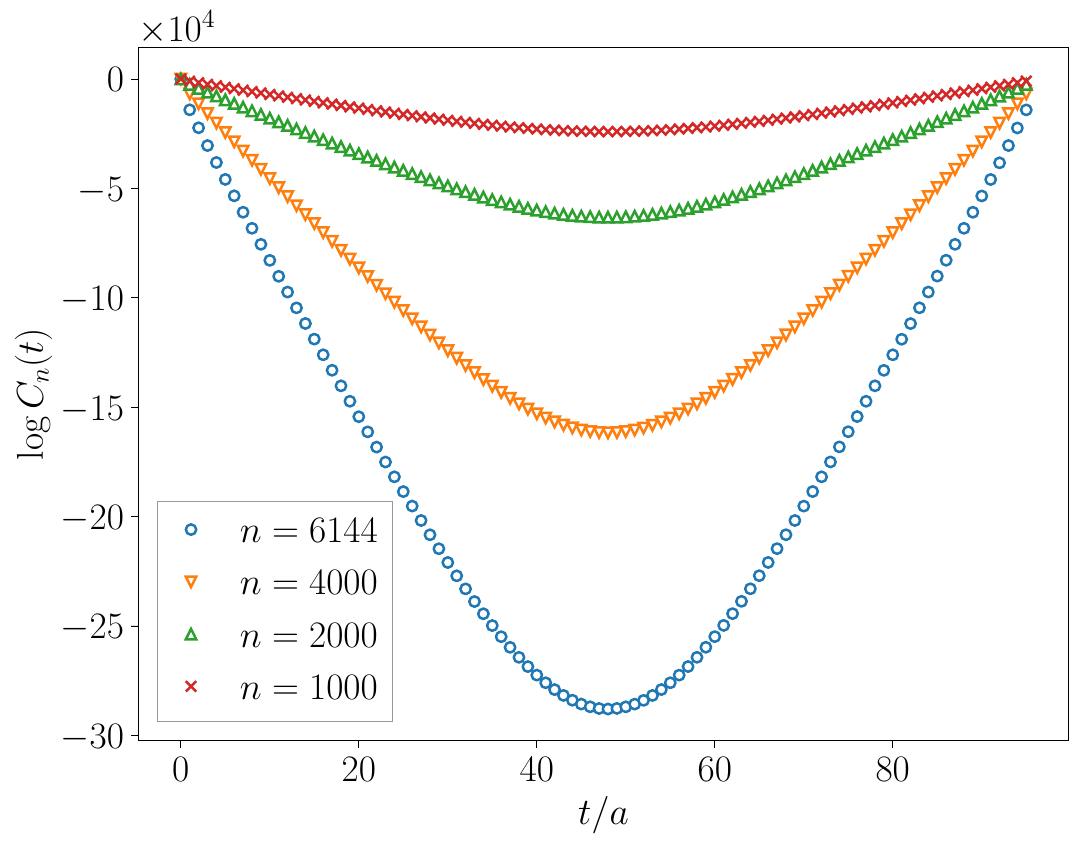}
    \caption{Correlation functions for $n \in \{1000$, 2000, 4000, 6144\} on a single configuration from the A ensemble.}
    \label{fig:single-config-correlators}
\end{figure}

\label{sec:lattice}

\subsection{Lattice Details}

All of the calculations in this work were performed on two ensembles, referred to as ensemble A and ensemble B, of gauge field configurations generated with Wilson-clover fermions and a tree-level tadpole-improved Symanzik gauge action, the parameters of which are summarized in Table~\ref{tab:lattice-params}. On both ensembles, measurements were separated by 10 hybrid Monte-Carlo trajectories; further details regarding these configurations are given in Ref.~\cite{Yoon:2016jzj}. Using these ensembles, we computed sets of smeared-source, smeared-sink propagators from a regular sparse grid on a single timeslice, $\Lambda_3=\{{\bf x} \mid x_i\, {\rm mod}\ s=0\ \forall  i\}$ with $s=6$ on ensemble A and $s=8$ on ensemble B, both corresponding to $N_\Lambda=512$. The source and sink smearings were gauge-covariant Gaussian smearing with 35 steps with width parameter 3.0 \cite{Edwards:2004sx}. These propagators are sparsened \cite{Detmold:2019fbk} in that the output was only stored on a sparse sub-lattice of the original lattice geometry. In this case, the same sparsening factors were used as in the choice of source locations.
Due to the sparsening, the dimensionality of the generalized spin-color-spatial matrix is $N=4\times3\times512=6144$, enabling correlation functions up to $n=6144$ to be computed.

\begin{table*}[t]
    \centering
\begin{tabular}{cccccccccc}
\hline
Label & $N_\text{conf}$ & $\beta$ & $C_{SW}$
& $a m_{ud}$ & $a m_s$ & $L^3 \times T$ & $a$ (fm) & $M_\pi$ (MeV)   & $M_\pi L$ \\
\hline
     A & 201 & 6.3 & 1.20537 & -0.2416 & -0.2050
    & $48^3 \times 96$ & 0.091(1) & 166(2) & 3.7  \\
     B & 322 & 6.3 & 1.20537 & -0.2416 & -0.2050
    & $64^3 \times 128$ & 0.091(1) & 172(6) & 5.08  \\ \hline
 \end{tabular}
    \caption{Parameters of the gauge-field configurations used in this work.
    The first column lists the label used to refer to the ensemble, $N_{\rm conf}$ is the number of configurations, and $\beta$ and $C_{\rm SW}$ refer to the gauge coupling and clover coefficient, respectively. The lattice spacing $a$ is determined in Ref.~\cite{Yoon:2016jzj}, while the lattice geometries are defined by the the spatial and temporal extents, $L$ and $T$, respectively. The bare light ($m_{ud}$) and strange ($m_s$) quark masses  are given in lattice units and $M_\pi$ is the pion mass determined in Ref.~\cite{Yoon:2016jzj}.}
    \label{tab:lattice-params}
\end{table*}

Except where otherwise stated, all of the calculations described below were performed at double precision. Calculations in double-double and triple-double precision show agreement with these to at least 1 part in $10^5$, as discussed in Appendix~\ref{sec:numeric-checks}.

\subsection{Single-configuration correlation functions}

\begin{figure}[t!]
    \centering
    \includegraphics[width=\linewidth]{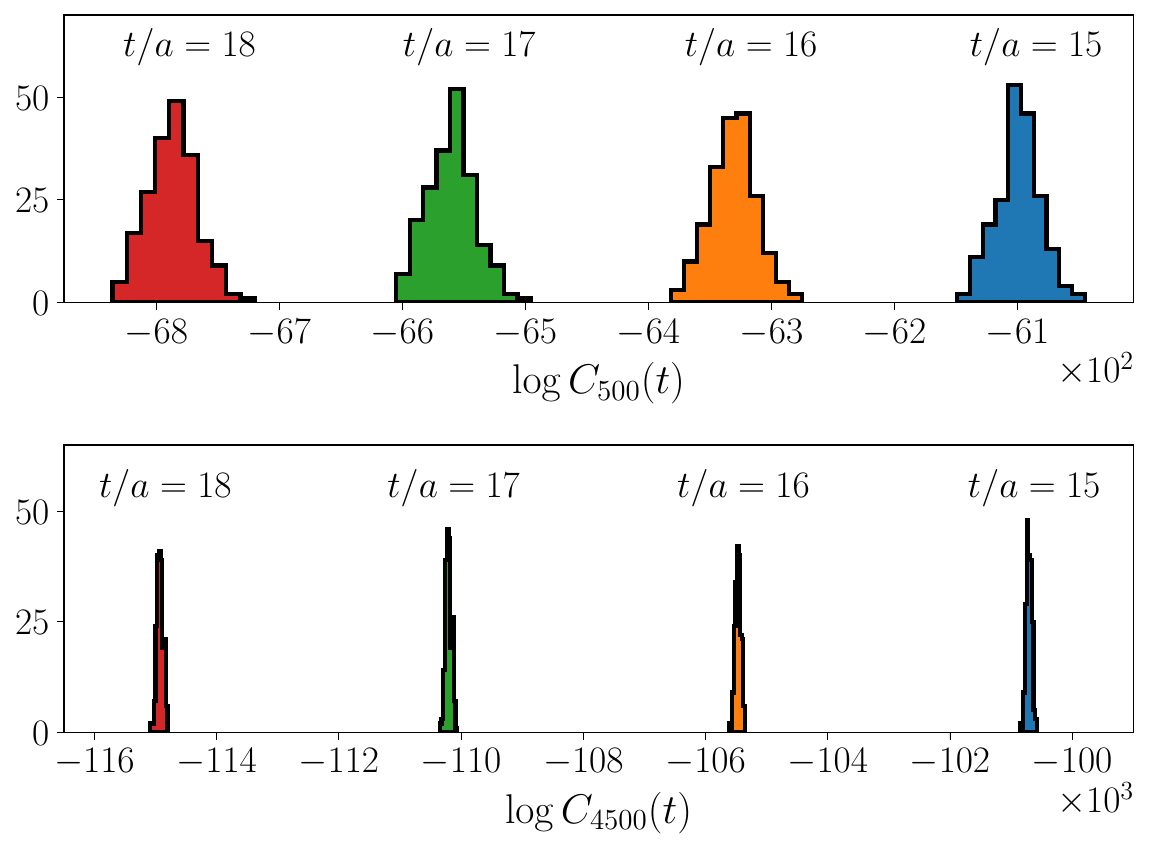}
    \caption{Histograms of the logarithms of correlation functions for $n\in\{500,4500\}$ at $t/a\in\{15, 16, 17,  18\}$ computed on the A ensemble.}
    \label{fig:histogram}
\end{figure}

In order to compute the pion correlation functions on each configuration, we first assembled the propagators into the spin-color-spatial matrix $S$, and then performed a SVD of $S$ as described in Sec. ~\ref{sec:calculating-eigenvalues}. We then combined the eigenvalues to form the pion correlation functions $C_n$ for $n \in \{ 1, \dots, 6144\}$ using the method described in Sec.~\ref{sec:sym-poly-algorithm}. Examples of the resulting correlation functions on a single configuration from the A ensemble are shown in Fig.~\ref{fig:single-config-correlators}. Notable here is the large variation in the scale of the correlation functions. For example, $C_{6144}(t)$ shown in Fig.~\ref{fig:single-config-correlators} ranges over more than $10^{5}$ orders of magnitude. Even on an individual timeslice, the correlation functions corresponding to a given number of pions evaluated on different configurations can vary by many orders of magnitude. This can be seen in Fig.~\ref{fig:histogram}, where we show histograms of the logarithms of correlation functions for a few adjacent timeslices for all 201 configurations of the A ensemble for $n\in\{500,4500\}$. Although the intra-timeslice variation seen in Fig.~\ref{fig:histogram} is small compared the inter-timeslice variation, it is still large enough to require special techniques to be used in analysing the correlation functions, as we will discuss in Sec.~\ref{sec:analysis}. Similar distributions are seen on the B ensemble.

\subsection{Distribution of eigenvalues}
\begin{figure}[t!]
    \centering
    \includegraphics[width=\linewidth]{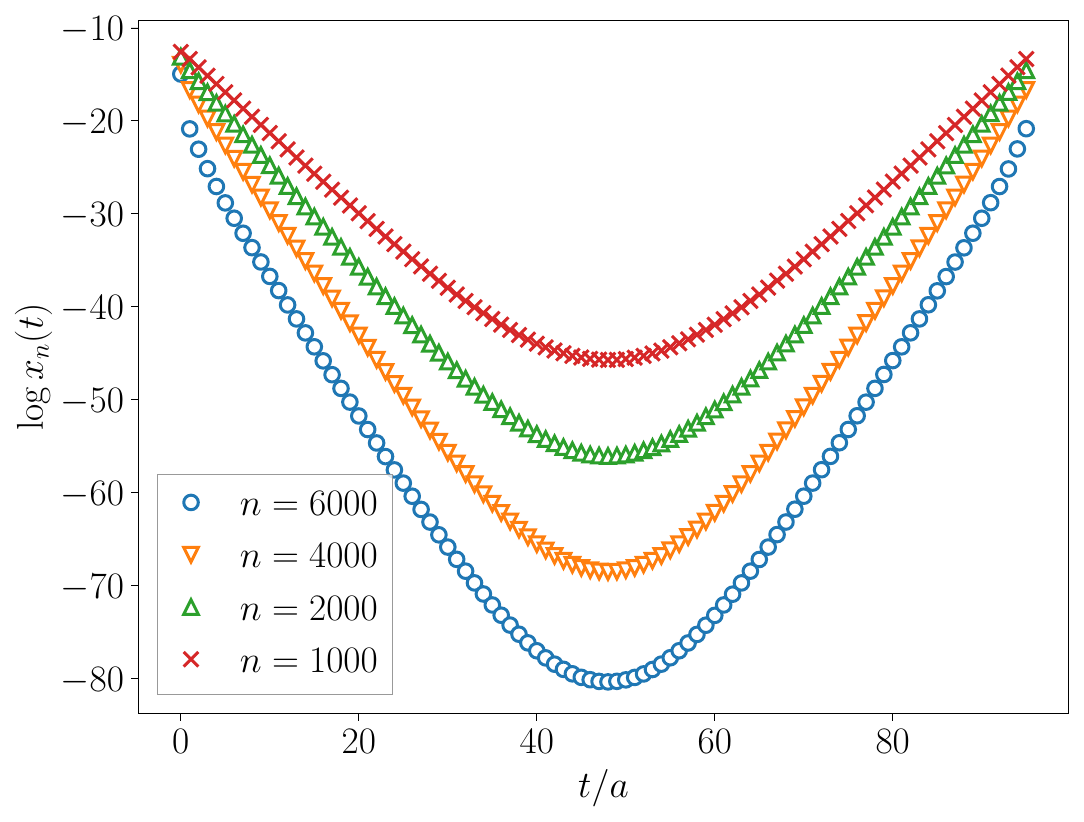}
    \caption{Logarithms of eigenvalues $x_n$ for $n\in\{1000,\ 2000,\ 4000,\ 6000\}$ as a function of timeslice on a single configuration from the A ensemble. Eigenvalues were computed using double-double precision.}
    \label{fig:eigenvalue-dist}
\end{figure}

While the eigenvalues of the pion block, $\vec x$, are not themselves physical, they are still of interest since they directly determine the correlation functions via Eq.~\eqref{eq:correlator-formula}. Values of $x_n$ for various choices of $n$ on a single configuration of the A ensemble over the full temporal extent of the lattice geometry are shown in Fig.~\ref{fig:eigenvalue-dist} (the B ensemble shows similar behavior).
Interestingly, the behaviour of each of the eigenvalues appears quite similar to that of an (ensemble-averaged) correlation function, exhibiting an exponential decay for moderate $t$. This behaviour is not physical as the eigenvalues are single-configuration quantities, however the exponential decay of the eigenvalues does have physical implications.
In particular, suppose that the exponential behaviour in Fig.~\ref{fig:eigenvalue-dist} were exact, so the $k^{\rm th}$ eigenvalue $x_k$ has the time dependence
\begin{equation}
\label{eq:eigenvalue-decay}
    x_k(t) = A_k \left[e^{-\alpha_k t}+ e^{-\alpha_k (T-t)}\right],
\end{equation}
where $A_k$ and $\alpha_k$ are constants. Then, combining these eigenvalues into a  function $\tilde C_n(t)$ via Eq.~\eqref{eq:correlator-formula}, and taking the limit $T \to \infty$ for simplicity, yields
\begin{equation}
    \tilde C_n(t) = n! \sum_{k_1 < \dots < k_n} A_{k_1} \dots A_{k_n}
    e^{-(\alpha_{k_1} + \dots + \alpha_{k_n}) t}.
\end{equation}
In this representation, $\tilde C_n(t)$ naturally behaves as a sum of exponentials, from which we can read off the energies
\begin{equation}
    \label{eq:eigenvalue-energy-sum}
    \tilde E_{k_1, \dots, k_n} = \alpha_{k_1} + \dots + \alpha_{k_n}; \hspace{0.5cm} \{k_i\} \text{ distinct.}
\end{equation}
This formula also describes the possible energies of an $n$-particle system of non-interacting fermions with single-particle energies $\{\alpha_k\}$. In an imprecise way, the rate of exponential decay for the eigenvalue $x_k(t)$ can be interpreted as corresponding to the $k^{\rm th}$-lowest single-particle energy for a pion in a volume $(aL)^3$. Note that this correspondence is not exact -- the quantity $x_k(t)$ is computed only on a single configuration, and the validity of Eq.~\eqref{eq:eigenvalue-decay} for describing $x_k(t)$ is empirical with no theoretical justification.

\section{Analyzing many-pion systems}

\label{sec:analysis}

\subsection{Central limit theorem-based methods}
Typical methods for analysing correlation functions in LQCD begin by collecting $N_{\rm conf}$ samples, $C_n^{[U_i]}(t)$, on independent\footnote{Throughout this work, we assume that the samples are sufficiently decorrelated to be effectively independent. Autocorrelations are seen to be small for the pion mass.} gauge-field configurations, $U_1, \dots, U_{N_\text{conf}}$, and computing the sample mean
\begin{equation}
\label{eq:correlator-empirical-average}
    \bar{C}_n(t) = \frac{1}{N_\text{conf}} \sum_{i = 1}^{N_\text{conf}} C_n^{[U_i]}(t),
\end{equation}
along with the sample variance
\begin{equation}
\label{eq:correlator-empirical-std}
 \Delta C_n^2(t) = \frac{1}{N_\text{conf}-1} \sum_{i = 1}^{N_\text{conf}} \left[ C_n^{[U_i]}(t) - \bar{C}_n(t)\right]^2.
\end{equation}
As $N_\text{conf} \to \infty$, the CLT applies and we may treat $\bar{C}_n$ as a Gaussian random variable with standard deviation $\sqrt{\Delta C_n^2/ N_\text{conf}}$. Performing correlated fits then allows the extraction of energies and other physical parameters of interest. In principle, these methods could be applied to correlation functions of the many-pion systems considered here. However, in practice the large range of scales involved in a many-pion correlation function makes analyses based on the CLT effectively impossible. In particular, as discussed above, the correlation functions on a particular timeslice can vary by many orders of magnitude, typically resulting in a single gauge configuration dominating the sample mean, Eq.~\eqref{eq:correlator-empirical-average}, far from the regime of applicability of the CLT.\footnote{In particular, this excludes the standard method of fitting the correlation function to a linear combination of exponential functions, as that method relies on the validity of the CLT.}

\begin{figure}[t!]
    \centering
    \vspace{-0.1cm}
    \includegraphics[width=\linewidth]{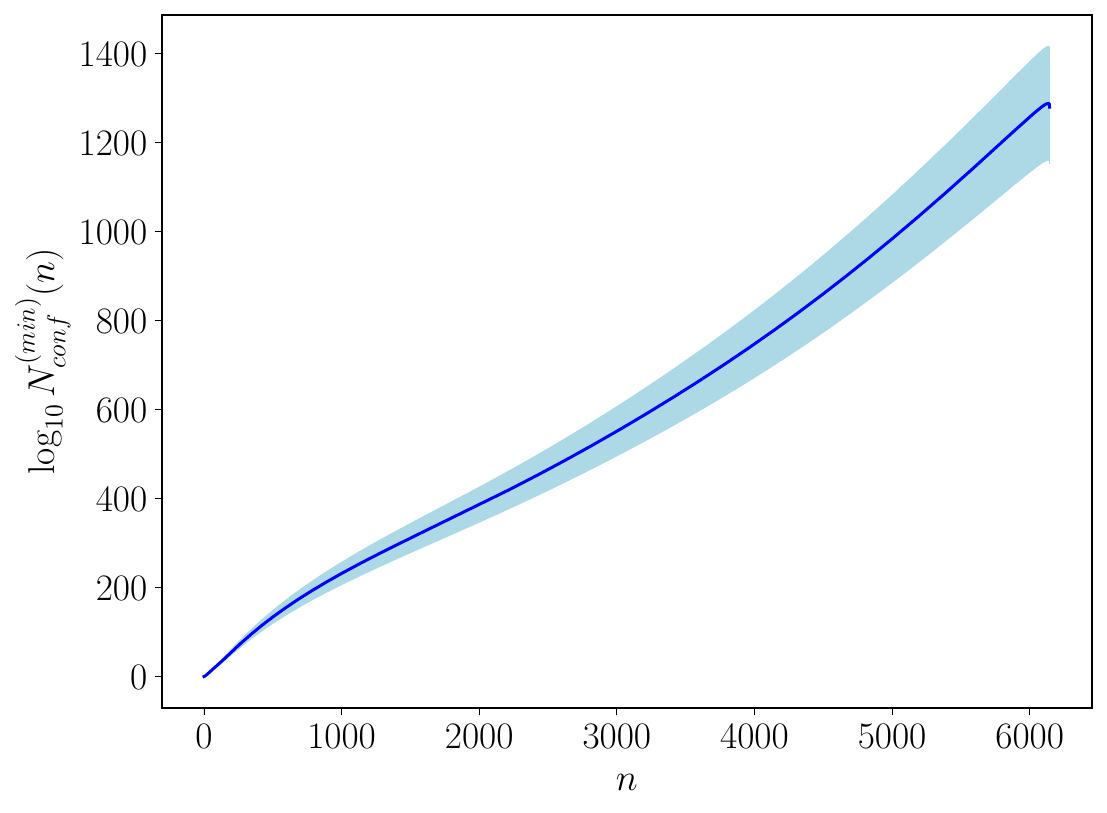}
    \caption{Estimate of the logarithm of the number of samples needed for the CLT to apply to correlation function $C_n(t)$ as a function of $n$. The uncertainty band indicates the standard deviation over bootstrap samples (see Sec.~\ref{sec:log-norm-analysis}).}
    \label{fig:digits-needed1}
\end{figure}

This argument can be made more precise. Suppose that that correlation functions $ C_n^{[U]}(t)$ were log-normally distributed across gauge configurations, i.e., for a given choice of $n$ and $t$, $\log  C_n^{[U]}(t) \sim \mathcal{N}(\mu_n, \sigma_n^2)$ is a normal distribution for some $\mu_n, \sigma_n$ (empirically, we observe that the sampled correlation functions are consistent with this assumption, as will be discussed below). Then using Eq.~(\ref{eq:correlator-empirical-average}), the expectation values $\braket{\bar{C}_n(t)}$ and $\braket{\bar{C}_n^2(t)}$ over the set of all ensembles can be determined to be
\begin{align}
    \braket{\bar{C}_n(t)} &= \exp \left( \mu_n + \frac{\sigma_n^2}{2} \right), \label{eq:correlator-log-normal-avg} \\
    \braket{\bar{C}_n^2(t)} &= \frac{1}{N_\text{conf}} \braket{\bar{C}_n(t)}^2 e^{\sigma_n^2}.
\end{align}
In order to satisfy the requirement for the CLT, it is necessary at a minimum that $\braket{\bar{C}_n^2} - \braket{\bar{C}_n}^2 \lesssim \braket{\bar{C}_n}^2$, which implies that $N_\text{conf} \gtrsim N_\text{conf}^\text{(min)}(n) = e^{\sigma_n^2}/2$. %
In Fig.~\ref{fig:digits-needed1}, we show an estimate of $N_\text{conf}^\text{(min)}(n)$ using
\begin{equation}
\label{eq:log-Cn-mean}
    \mu_n = \frac{1}{N_\text{conf}}
    \sum_{i=1}^{N_{\text{conf}}} 
    \log C_n^{[U_i]}(t)
\end{equation}
and
\begin{equation}
\label{eq:log-Cn-var}
    \sigma_n^2 = \frac{1}{N_\text{conf} - 1}
    \sum_{i=1}^{N_{\text{conf}}} \left(
    \log C_n^{[U_i]}(t) - \mu_n
    \right)^2.
\end{equation}
As can be seen, $N_\text{conf}^\text{(min)}(n)$ grows rapidly and already for $n\sim 100$ is an unrealistically large number of configurations is needed, effectively ruling out the use of standard statistical methods for $n \gtrsim 100$ pion systems. Note that the value of $N_\text{conf}^\text{(min)}(n)$ will depend on the choice of quark masses and physical volume. 

\subsection{Log normality}
\label{sec:log-normal}
\begin{figure}[t!]
    \centering
    \vspace{-0.1cm}
    \includegraphics[width=\linewidth]{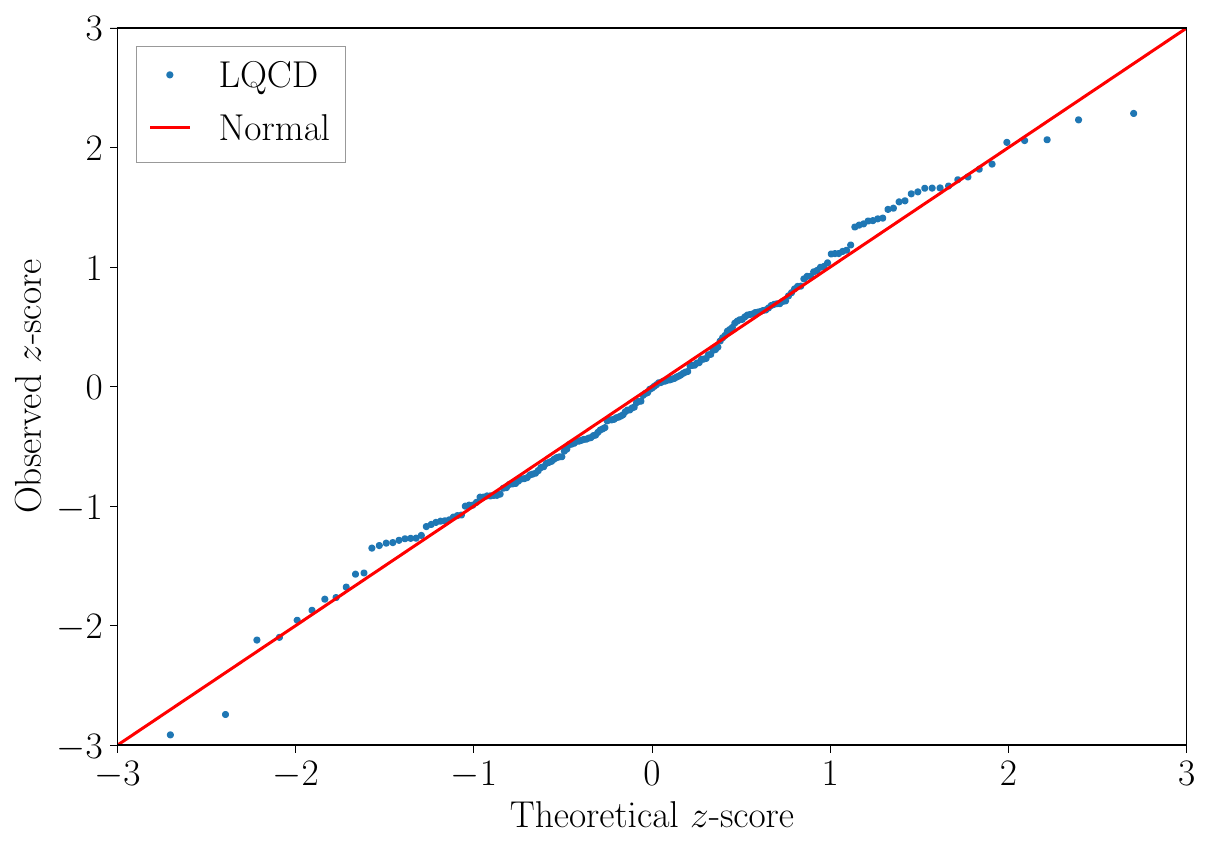}
    \caption{Quantile-quantile plot for the distribution of $\log C_{6000}(t=18a)$ on the A ensemble. Each point represents a different configuration $U_i$, with the vertical position indicating the $z$-score (number of standard deviations from the mean) of $\log C^{[U_i]}_{6000}(t=18a)$, while the horizontal position of the point indicates the theoretical $z$-score for the corresponding quantile of a normal distribution. The red line indicates the theoretical expectation for a normal distribution.}
    \label{fig:qqplot}
\end{figure}

\begin{figure}[t!]
    \centering
    \vspace{-0.1cm}
    \includegraphics[width=\linewidth]{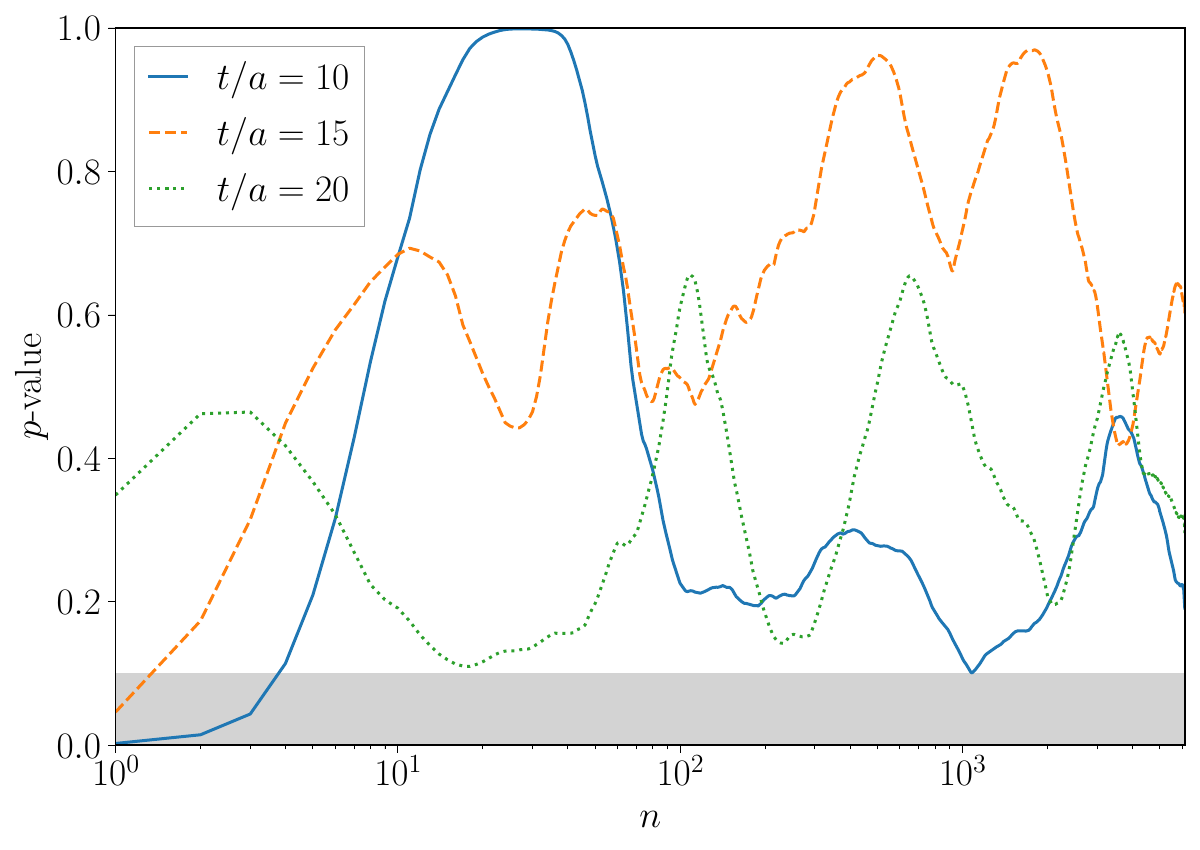}
    \caption{The Shapiro-Wilk test $p$-values as a function of $n$ at timeslices $t/a\in\{10,\ 15,\ 20\}$ for the A ensemble. A value of $p\lesssim0.1$ (gray band) indicates a violation of log-normality of the correlation-function distribution across configurations.}
    \label{fig:shapiro-wilk}
\end{figure}

Since CLT-based methods such as the sample mean are not applicable for many-pion correlation functions at the statistical precision achieved in this work, we need to use a different method of analysis. 
A path forward is provided by the data which, from a cursory inspection of the distributions in Fig.~\ref{fig:histogram}, appears to be approximately log-normal, as mentioned above. This observation can be also be seen qualitatively in Fig.~\ref{fig:qqplot}, which shows the observed quantiles of $\log C_n^{[U_i]}(t)$ on the A ensemble for particular choices of $n$ and $t$ against the expected quantiles for a normal distribution, showing the approximate log-normality of the samples. Quantile-quantile plots for other choices of $n$ and $t$ show similar behavior on both ensembles. To verify this observation of log-normality, we employ the Shapiro-Wilk test \cite{10.1093/biomet/52.3-4.591}, which is designed to assess whether a given set of samples, in this case the logarithms of the correlation functions, have a distribution consistent with  a normal distribution. The resulting $p$-values for the tests for different $n$ and $t$ are shown in Fig.~\ref{fig:shapiro-wilk} for the A ensemble (similar behavior is seen for the B ensemble). With the exception of the $(n<5)$-pion systems, none of the correlation function distributions in this study have a $p$-value less than 0.1, indicating that we do not observe violations of log-normality on these samples. Since we cannot detect statistical violations of log-normality, we conclude that any bias induced by the assumption of log-normality is likely subdominant to the statistical uncertainties of our estimates. This is not entirely unexpected -- log normal random variables often appear when taking products of many non-negative random variables (particular projections of propagators in this case), and it has been previously hypothesized that log-normality may play a role in QCD correlation 
functions~\cite{Guagnelli:1990jb,MJSPC,endres_noise_2011,DeGrand:2012ik}. 

Under the assumption that correlation functions are drawn from a log-normal distribution, i.e., $\log C_n^{[U]} \sim \mathcal{N}(\mu_n, \sigma_n^2)$, we can obtain a lower-variance estimator by determining the parameters $\mu_n$ and $\sigma_n^2$ via Eqs.~\eqref{eq:log-Cn-mean} and \eqref{eq:log-Cn-var} and then using the analytic form for $\braket{C_n}$ given in Eq.~\eqref{eq:correlator-log-normal-avg} to estimate the original correlation function. 
Should violations of log-normality be observed at higher statistical precision, it would be possible to systematically improve this method through the inclusion of higher cumulants, as discussed in \Cref{sec:cumulants}.

\subsection{Log-normal Analysis}\label{sec:log-norm-analysis}
\begin{figure}[t!]
    \centering
    \vspace{-0.1cm}
    \includegraphics[width=\linewidth]{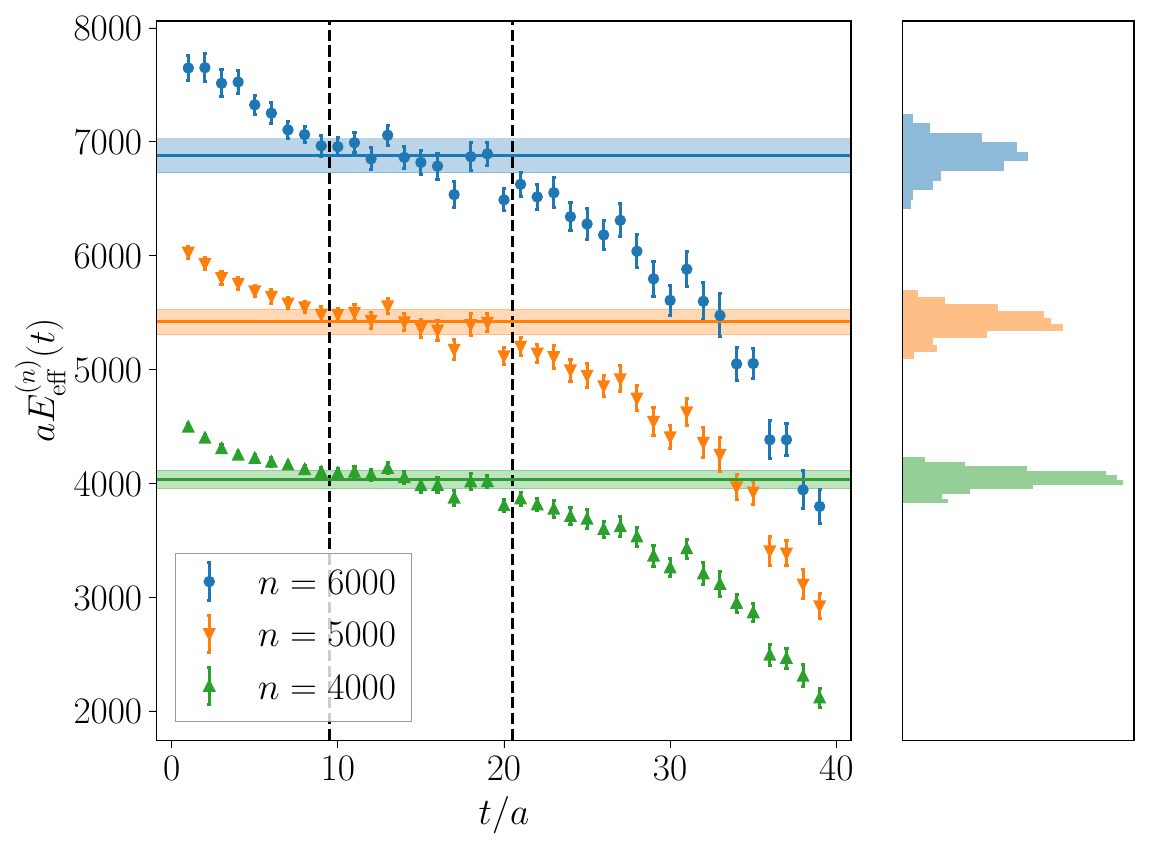}
    \caption{
    Effective energy functions calculated for ${n\in\{4000,\ 5000,\ 6000\}}$ on the A ensemble. The vertical extents of the shaded bands indicate the extracted fit energies, while the histograms in the right panel show the distributions of the energies across bootstrap samples. The vertical black dashed lines indicates the timeslices included in the procedure used to extract the energy, as discussed in the main text.}
    \label{fig:effective_mass}
\end{figure}

In order to extract energies from the computed correlation functions, we first produce a set of 200 bootstrap samples~\cite{bootstrap}, and then compute the mean $\mu_n(t)$ and standard deviation $\sigma_n(t)$ of $\log C_n^{[U]}(t)$ on each bootstrap sample. We then combine these quantities to form bootstrap estimates of 
\begin{equation}
    C_n(t) = \exp \left( \mu_n(t) + \frac{\sigma_n(t)^2}{2} \right),
\end{equation}
and the effective energy defined by
\begin{equation}
    \begin{aligned}
    E_\text{eff}^{(n)}(t) = &\log \frac{C_n(t)}{C_n(t - 1)} \\
    = &\ \mu_n(t) - \mu_n(t-1) + \frac{\sigma_n^2(t)}{2}
    -  \frac{\sigma_n^2(t-1)}{2} ,
\end{aligned}
\end{equation}
which asymptotes to the ground-state energy for asymptotic $t$ and lattice temporal extent.
Examples of the effective energies are shown in Fig.~\ref{fig:effective_mass}. The uncertainties are quantified using the the standard deviation over bootstrap samples. All uncertainties on LQCD quantities shown below indicate the standard deviation over the bootstrap samples.

\begin{figure}[t!]
    \centering
    \vspace{-0.1cm}
    \includegraphics[width=0.98\linewidth]{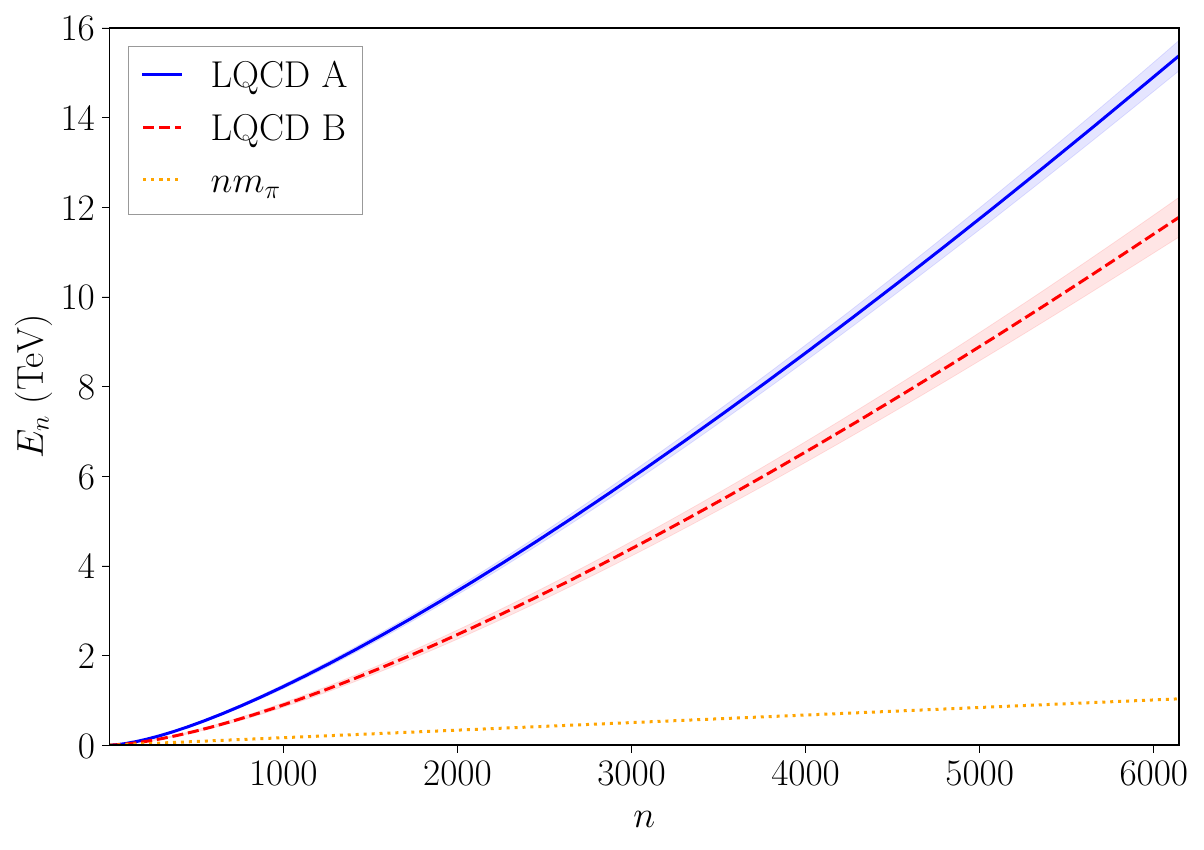}
    \caption{Energies of the multi-pion systems as a function of $n$ on both the A ($48^3\times96$) and B ($64^3\times 128$) ensembles, labeled ``LQCD A'' and ``LQCD B'', respectively.
    The shaded bands represent the uncertainty as calculated from the variance over the bootstrap results. }
    \label{fig:effective_mass_all}
\end{figure}

As can be seen from the effective energy functions, the correlation functions are contaminated by both excited states at early times and by thermal effects near the middle of the lattice temporal extent. Determining the ground-state energy for each $n$ from these signals is challenging because the excited-state and thermal effects are not small and there are significant statistical fluctuations within the time range in which the signal is consistent with a constant.
In order to take a conservative approach to energy extraction, on each bootstrap sample, we take the effective mass from a single timeslice drawn from the uniform distribution over
$t/a \in [10,20] \cup [76,86]$.
This encompasses a variety of different fitting choices and ensures that the energy uncertainty represents an envelope over different fit procedures as well as statistical fluctuations. Figure~\ref{fig:effective_mass} shows the resulting fitted values and uncertainties for three different values of $n$ for the A ensemble. We find that the uncertainty band on the fitted energy is compatible with the distribution of the effective energies within the region of the fit.\footnote{Here we refer to a set of data points $x_i$ with associated uncertainties $\sigma_i$ as compatible with a fit $x_\text{fit}$ with uncertainty $\sigma_\text{fit}$ if the average of $(x_i - x_\text{fit})^2 / (\sigma_i^2+\sigma_\text{fit}^2)$ is $\lesssim 1$.}
The correlation functions on ensemble B have a larger temporal extent, so fits are performed in the interval $t/a \in [10,25] \cup [103,118]$; the fits to extract the energies display similar behavior as on ensemble A. 
The $n$ dependence of the extracted energies $E_n$ on both ensembles is shown in Fig.~\ref{fig:effective_mass_all}.  There are strong correlations between correlation functions for different $n$ that will be exploited below.

\section{Large isospin chemical potential}
\label{sec:isospin}
\begin{figure}[t!]
    \centering
    \vspace{-0.1cm}
    \includegraphics[width=\linewidth]{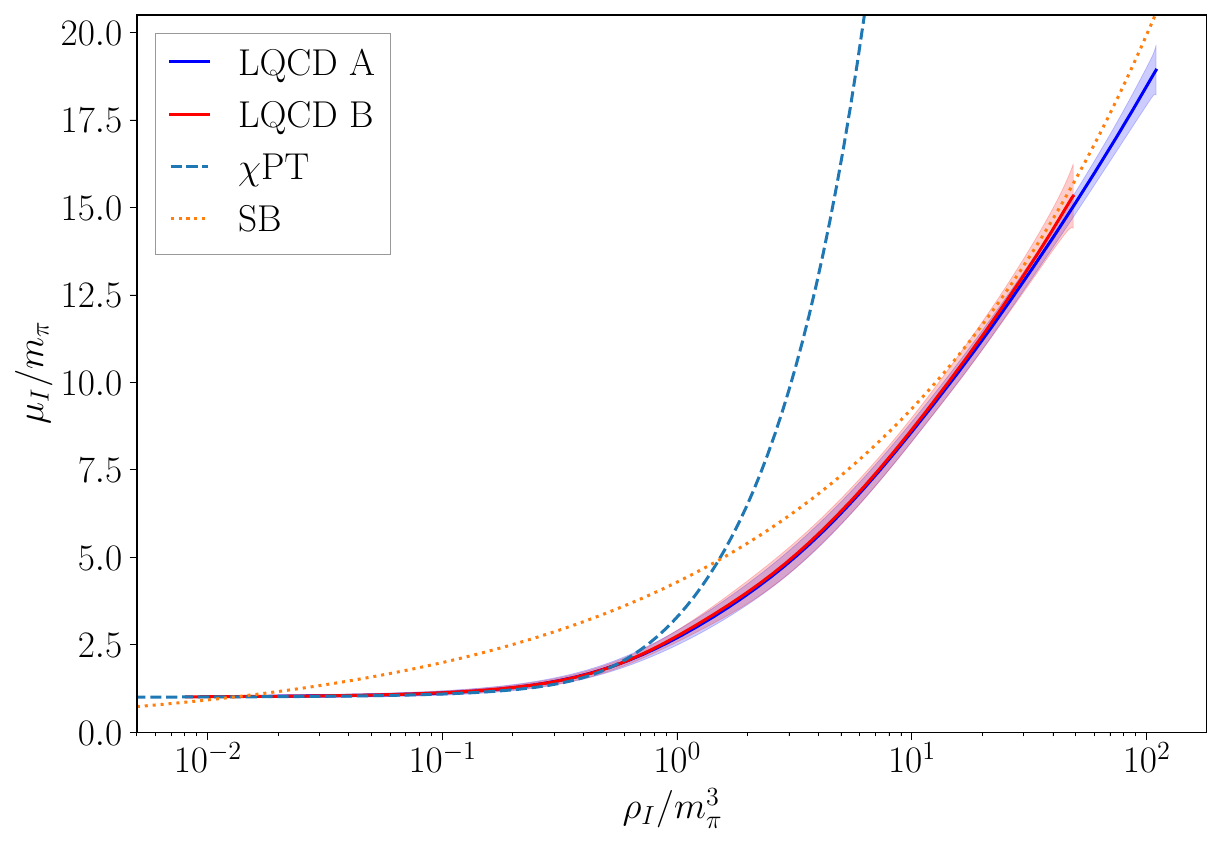}
    \caption{The isospin chemical potential of the many-pion systems studied in this work as a function of the isospin density for both ensemble A and ensemble B. Error bands are obtained from the standard deviation over bootstrap samples. For comparison, the expectations from $\chi$PT and Stefan-Boltzmann (SB) limit as blue-dashed and orange-dotted lines, respectively.}
    \label{fig:chemical-potential}
\end{figure}

\begin{figure}[t!]
    \centering
    \vspace{-0.1cm}
    \includegraphics[width=\linewidth]{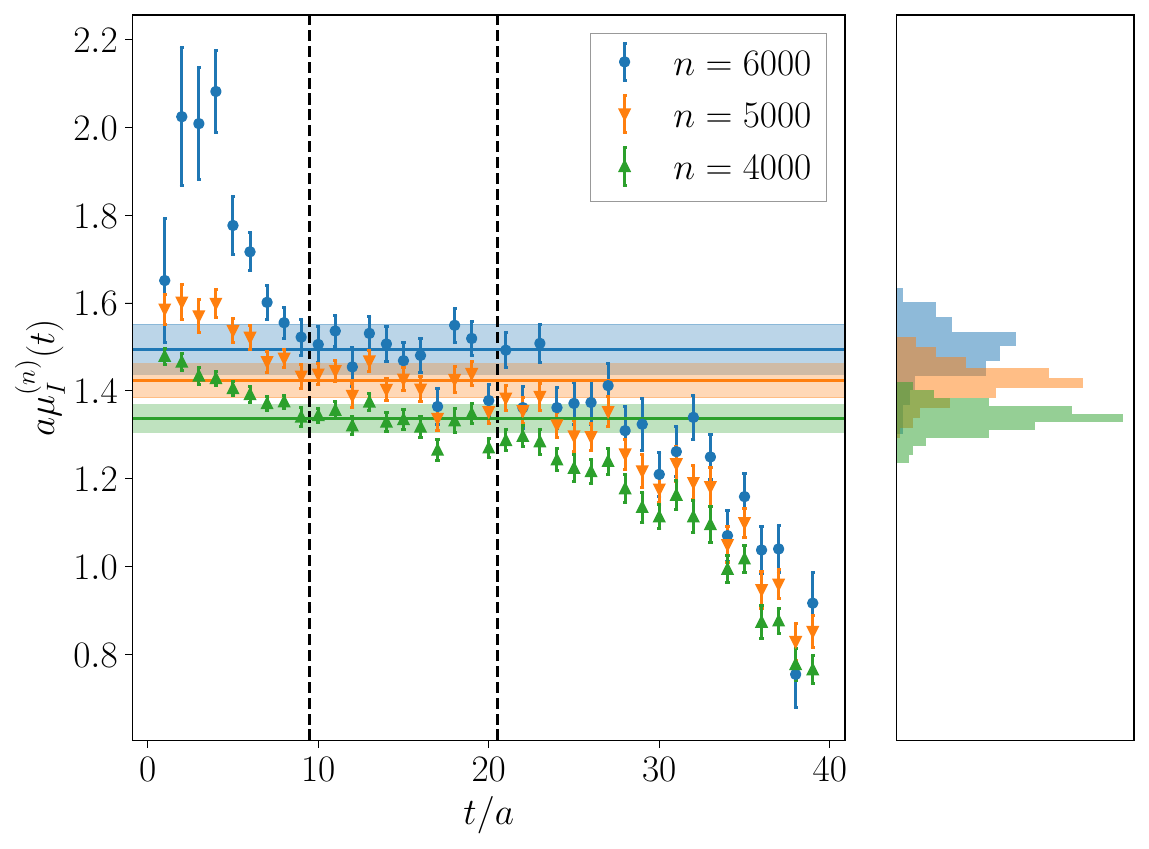}
    \caption{The effective chemical-potential function $\mu_{I}^{(n)}(t)=(E_{\rm eff}^{(n+1)}(t)-E_{\rm eff}^{(n-1)}(t))/2$ as a function of the temporal separation used for $n\in\{4000,\,5000,\,6000\}$ on the A ensemble. The vertical extent of the shaded bands indicates the uncertainty in the chemical potential, and the histograms in the right panel show the distributions of the bootstrap samples. The black dashed vertical lines indicate the temporal extent included within the procedure used to determine $\mu_I$, as discussed in the main text.}
    \label{fig:chemical-potential-time-dependence}
\end{figure}

The isospin chemical potential of a system with a $z$-component of isospin $I_z=n$ and volume $V$ is defined as
\begin{equation}
    \mu_I(n) = \left. \frac{\mathrm{d} E_n}{\mathrm{d} n} \right|_{V = \text{const}}.
\end{equation}
Given a set of energies $\{E_n\}$ for $n$-$\pi^+$ systems in a fixed volume $V$, the isospin chemical potential $\mu_I(\rho_n)$ at density $\rho_I = \rho_n = n / V$ can be estimated via a finite-difference approximation\footnote{Higher-order stencils for the finite difference lead to results that are indistinguishable within the uncertainties.}
\begin{equation}
    \mu_I(\rho_n) = \frac{E_{n+1} - E_{n-1}}{2}.
\end{equation}
Using the bootstrap values of $E_n$ determined above, the resulting isospin chemical potentials on the two ensembles are shown as a function of the isospin density in Fig.~\ref{fig:chemical-potential}.\footnote{In Fig.~\ref{fig:chemical-potential} and all further figures, only values up to $n=6000$ are shown due to the large uncertainties for $n>6000$.} The dependence of the extracted chemical potential on the chosen temporal separation is shown in Fig.~\ref{fig:chemical-potential-time-dependence}; the correlations between energies for neighbouring values of $n$ result in the chemical potential being determined orders of magnitude more precisely than the individual energies. Notably, the results from both ensembles are consistent within uncertainties. Similar agreement is found in all of the observables shown below, indicating that finite-volume and finite-temperature effects are small and both lattice calculations are near the thermodynamic limit.

These results are compared with two predictions in Fig.~\ref{fig:chemical-potential}.
First, a result derived from leading order chiral perturbation 
theory ($\chi$PT) \cite{son_qcd_2001,Carignano:2016rvs} is that\footnote{Here, we use the convention in which the pion decay constant is $f_\pi \sim 132 \, \mathrm{MeV}$ at the physical values of the quark masses, as in Ref~\cite{Detmold:2008fn}.}
\begin{equation}
    \rho_I = \frac{1}{2} f_{\pi}^2 \mu_I
    \left( 1 - \frac{m_\pi^4}{\mu_I^4} \right).
\end{equation}
This relation is expected to be valid for low density systems in which the pions are weakly interacting but will break down as the isospin density or chemical potential becomes large compared to the chiral symmetry breaking scale. The second model is that of a relativistic fermion gas in the Stefan-Boltzmann (SB) limit, in which
\begin{equation}
\label{eq:stefan-boltzmann-muI}
    \mu_I = \left(\frac{48 \pi^2 \rho_I}{N_f N_c} \right)^{1/3},
\end{equation}
where $N_c \times N_f$ degrees of freedom are assumed with $N_f=2$ and $N_c=3$.
Notably, the Stefan-Boltzmann prediction does not have any free parameters, so the qualitative agreement between the LQCD data and the prediction in Eq.~\eqref{eq:stefan-boltzmann-muI} is somewhat remarkable and is quite suggestive as to the nature of the high-density state.

\begin{figure}[t!]
    \centering
    \vspace{-0.1cm}
    \includegraphics[width=\linewidth]{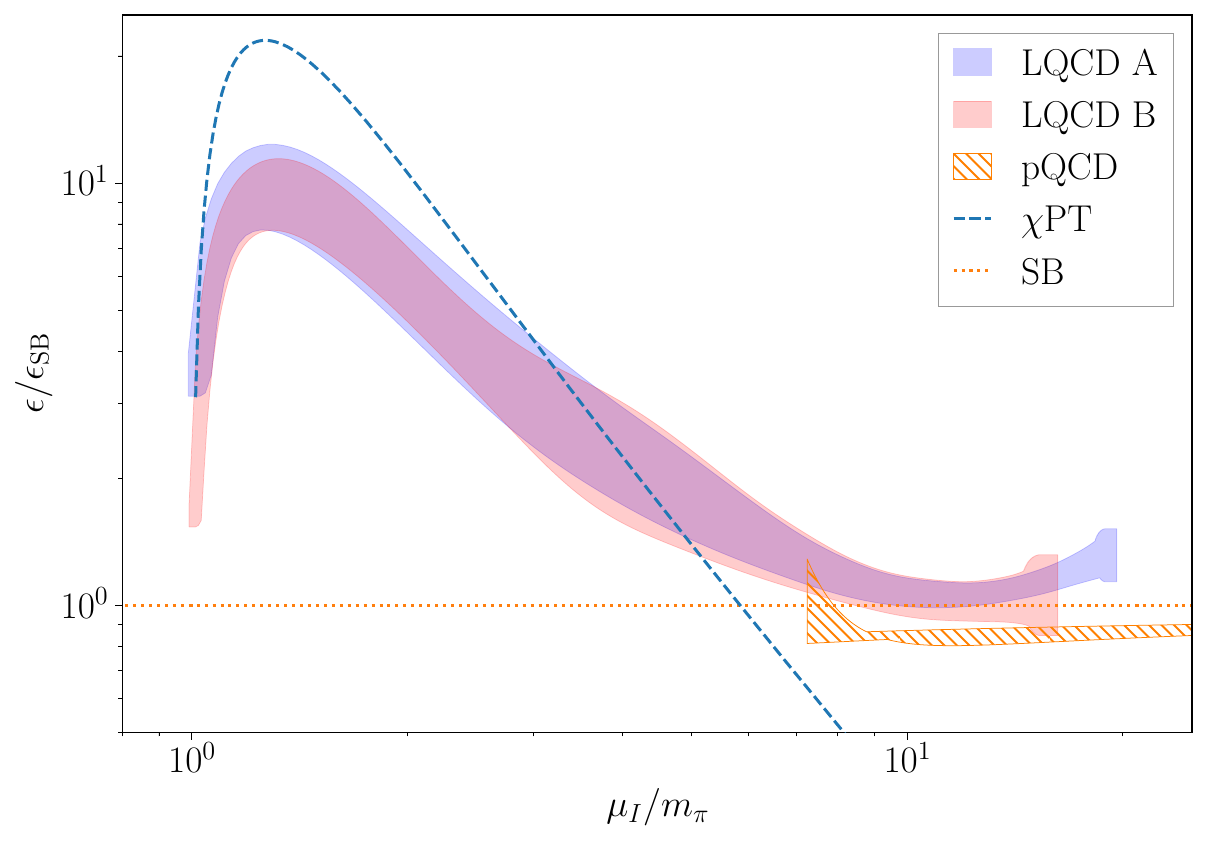}
    \caption{ The ratio of the energy density of the many-pion systems to the Stefan-Boltzmann prediction, Eq.~\eqref{eq:stefan-boltzmann-epsilon}, for the A and B lattice ensembles. The blue (A) and red (B) shaded regions represent interpolations of the LQCD results and their uncertainties as discussed in Appendix \ref{app:plotting}. Also shown are expectations from chiral perturbation theory (blue dashed line) and perturbative QCD at next-to-leading order (NLO)~\cite{Graf:2015tda} (orange hatched region). The uncertainties on the perturbative QCD result are obtained by varying the renormalization scale $\Lambda$ between $\mu_I/4$ and $\mu_I$.}
    \label{fig:energy-density-ratio}
\end{figure}

In Fig.~\ref{fig:energy-density-ratio}, we show the energy density $\epsilon_n=E_n/L^3$ as a function of the corresponding isospin chemical potential, normalized to the Stefan-Boltzmann expectation. In this and subsequent figures, we show an interpolation of the ${\cal O}(6000)$ discrete LQCD data points for each ensemble, using the approach presented in Appendix \ref{app:plotting} to produce a region that represents the horizontal and vertical uncertainties in the data.
For large $\mu_I$, the energy density is expected to match that of a $N_f = 2$, $N_c=3$ flavor fermion gas, namely
\begin{equation}
\label{eq:stefan-boltzmann-epsilon}
    \epsilon_{\text{SB}} = \frac{N_f N_c}{4 \pi^2} \left(\frac{\mu_I}{2}\right)^4.
\end{equation}
For comparison, we also show predictions from $\chi$PT~\cite{son_qcd_2001,Carignano:2016rvs} and one-loop perturbative QCD~\cite{Graf:2015tda}. Notably, these predictions agree qualitatively with the LQCD results in their respective regions of validity, namely small $\mu_I$ for $\chi $PT and large $\mu_I$ for perturbative QCD.\footnote{Predictions for thermodynamic quantities at nonzero isospin chemical potential can also be made in the Nambu--Jona-Lasinio  model~\cite{Nambu:1960xd,Nambu:1961tp,Nambu:1961fr} whose parameters  can be tuned in such a way that its predictions agree with $\chi $PT and LQCD in the low $\mu_I$ region, as shown in Refs.~\cite{He:2005nk,Xia:2013caa,Mannarelli:2019hgn,Lopes:2021tro}.} 
For very large $\mu_I$, there is a slight discrepancy between the LQCD results on ensemble A and the perturbative QCD expectation; however, given that the systematic uncertainties from discretization effects are not controlled in this study, it is unclear whether the LQCD results at the largest $\mu_I$  are reliable.
On dimensional grounds, lattice artifacts are expected to be suppressed by powers of the quark chemical potential, $\mu_I a/2$, which reaches 0.7 for the largest isospin chemical potential that is considered. Excited-state contamination is also not well controlled in the energy fits. Further exploration with calculations at a smaller lattice spacing, larger temporal extents, and higher statistical precision is needed to investigate these effects. Nonetheless, viewed globally, the LQCD data agree qualitatively with both low- and high-density expectations, smoothly interpolating between the two regimes. 

\begin{figure}[t!]
    \centering
    \vspace{-0.1cm}
    \includegraphics[width=\linewidth]{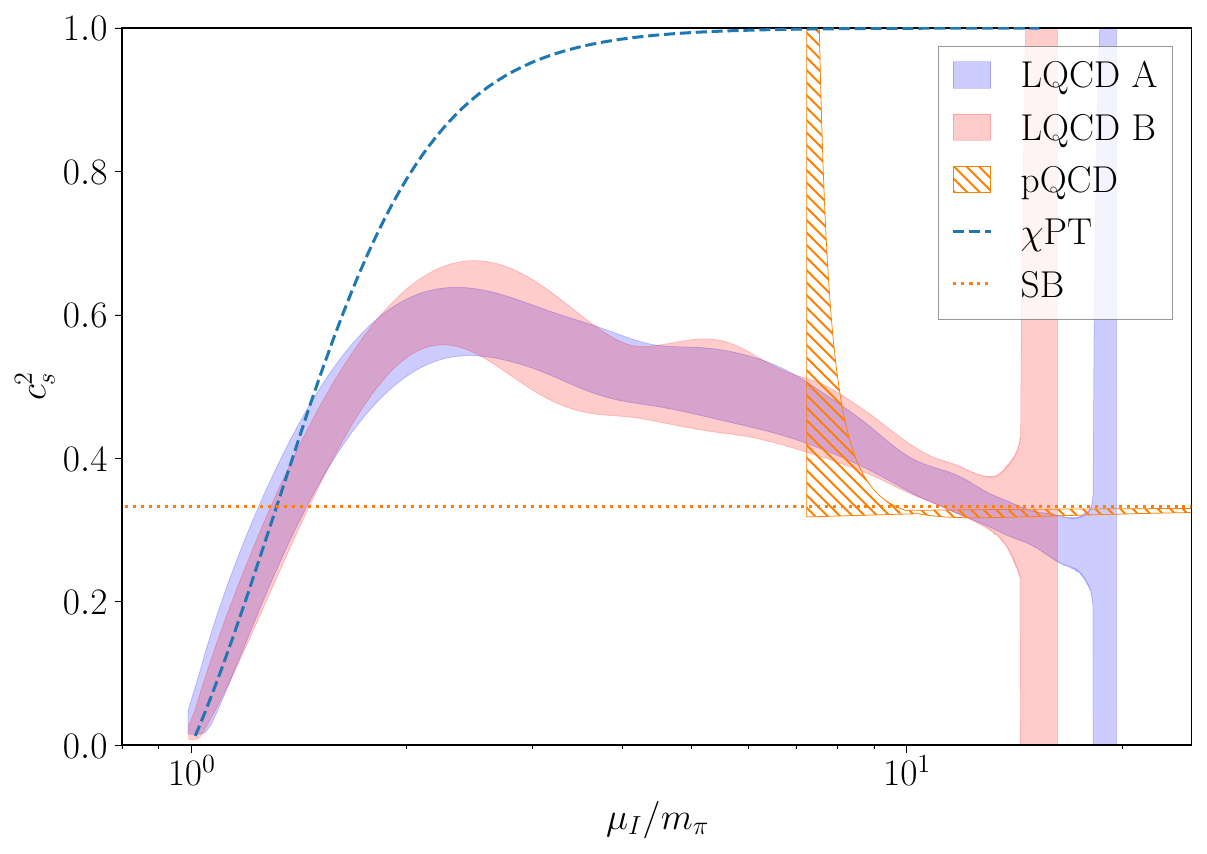}
    \caption{The squared speed of sound computed as in Eq.~\eqref{eq:speed-of-sound} as a function of the isospin chemical potential on ensemble A (blue) and ensemble B (red). The expectations in perturbative QCD (orange hatched region), chiral perturbation theory (blue dashed curve) and the Stefan-Boltzmann limit  (orange dotted line) are shown for comparison.}
    \label{fig:speed-of-sound}
\end{figure}

From the chemical potential and energy density, additional thermodynamic quantities characterizing high-isospin-density matter can be computed. 
A particularly important example is the speed of sound defined as (using units where the speed of light is $c=1$)
\begin{equation}\begin{aligned}
    c_s^2 &= \, \frac{dp}{d\epsilon} = \frac{n}{\mu_I}\frac{d\mu_I}{dn} = \frac{n}{dE/dn} \frac{d^2E}{dn^2}
    \\
    &\approx 2n \frac{E_{n + 1} - 2 E_n + E_{n-1}}{E_{n+1} - E_{n-1}}\,,
\end{aligned}
    \label{eq:speed-of-sound}
\end{equation}
where $p$ is the pressure.\footnote{The vacuum-subtracted pressure, $p$, is computed by numerically integrating the relation $\frac{dp}{dn}=\frac{n}{V}\frac{d \mu_I}{dn}$.} This governs isentropic propagation of sound waves through the medium (the isentropic condition is appropriate since our calculations correspond to a temperature that is close to zero, $T\sim 23$ MeV and 17 MeV for ensembles A and B, respectively).
The speed of sound is shown as a function of the isospin chemical potential in units of the pion mass in Fig.~\ref{fig:speed-of-sound} where it is seen to exceed the ideal gas limit. As for the energy density, close agreement is seen between the results from the two lattice ensembles. A similar result has been found in Ref.~\cite{Brandt:2022hwy}; however a larger range of $\mu_I/m_\pi$ is accessible in the current work. In particular, $c_s^2$ exceeds $1/3$ for $1.5\alt \mu_I/m_\pi\alt 14$, rising to a maximum of $c_{s,{\rm max}}^2\sim 0.6$ at $\mu_I\sim 2 m_\pi$ before decreasing back to the ideal-gas limit for large $\mu_I$. A maximum speed of sound above the ideal-gas limit  at intermediate values of chemical potential is also seen in two-color QCD \cite{Iida:2022hyy} and quarkyonic models \cite{McLerran:2018hbz}, but is in contradiction to the predictions of leading-order chiral perturbation theory in which $c_s$ rises monotonically to 1. This behavior is indicative of additional degrees of freedom other than in-vacuum pions becoming excited in the medium. From the numerical results herein, it remains an open question as to whether the speed of sound approaches the free gas limit from below (as expected from  perturbation theory \cite{Graf:2015tda}) or from above (as expected from resummed perturbation theory \cite{Fujimoto:2020tjc} or from the inclusion of power corrections \cite{Chiba:2023ftg}). 

\begin{figure}[t!]
    \centering
    \vspace{-0.1cm}
    \includegraphics[width=\linewidth]{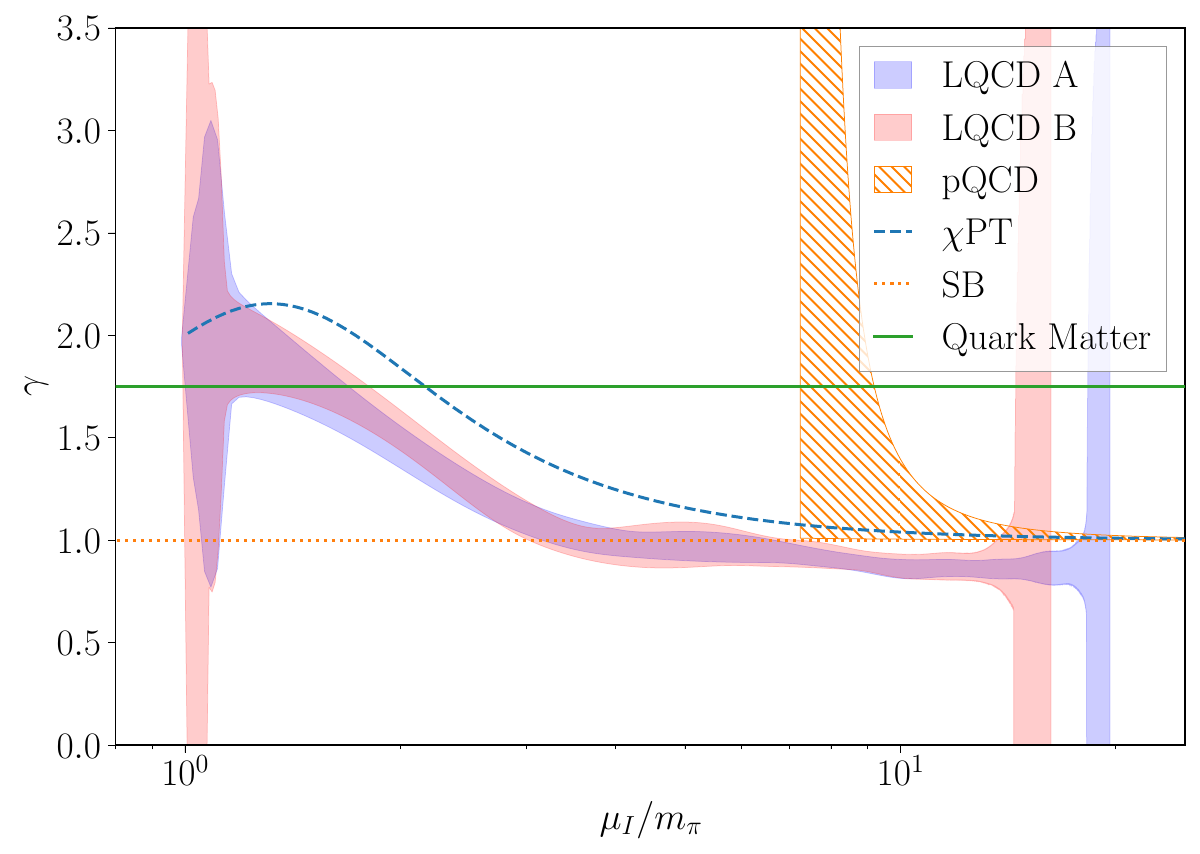}
    \caption{The polytropic index, $\gamma$, as a function of the isospin chemical potential on the A(B) ensemble is shown as the blue(red) region.
    The expectations in perturbative QCD (orange hatched region), chiral perturbation theory (blue dashed curve) and the Stefan-Boltzmann limit (orange dotted line) are shown for comparison. In addition, the bound at $\gamma=1.75$ below which the medium is expected to correspond to quark degrees of freedom \cite{Annala:2019puf} is indicated as the green horizontal line.}
    \label{fig:polytropic-index}
\end{figure}

\begin{figure}[t!]
    \centering
    \vspace{-0.1cm}
    \includegraphics[width=\linewidth]{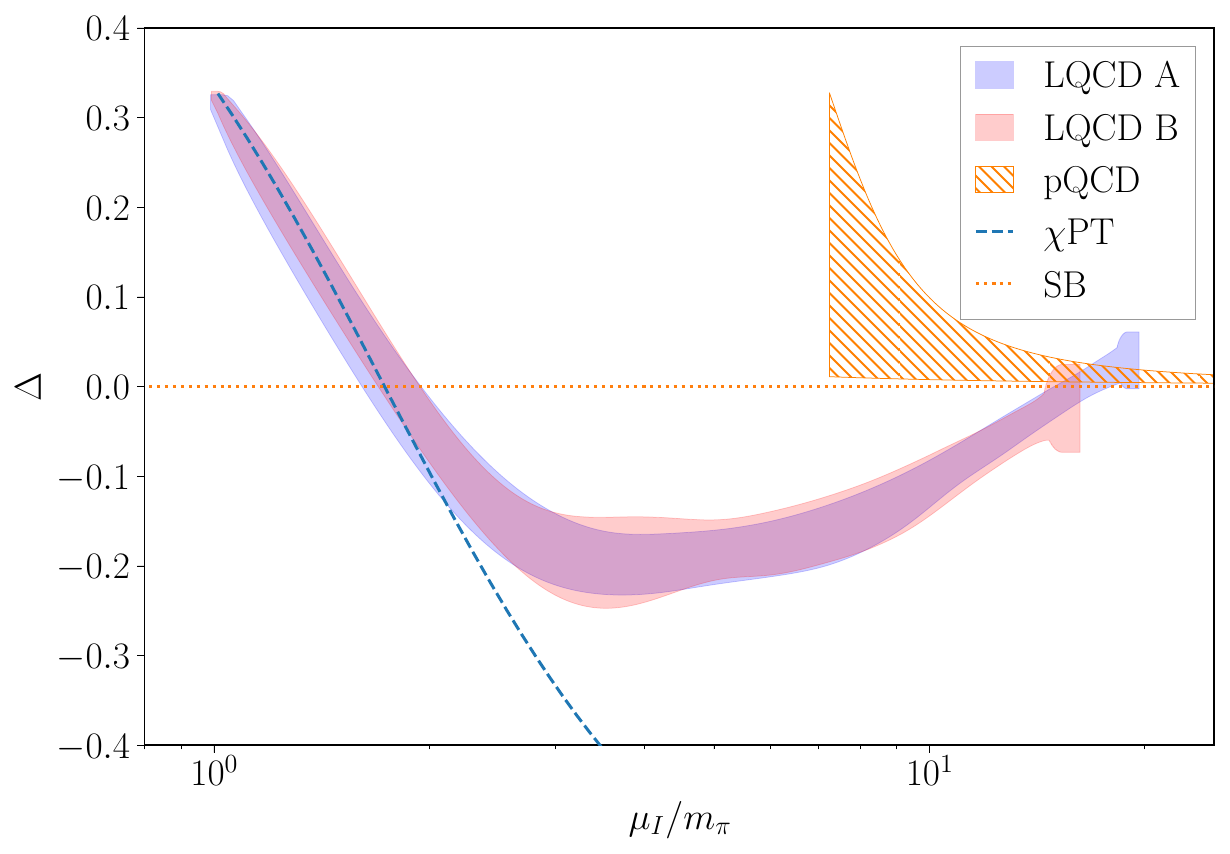}
    \caption{The normalized trace anomaly, $\Delta$,  as a function of the isospin chemical potential on the A(B) ensemble is shown as the blue(red) region. This quantity is bounded as $-2/3<\Delta < 1/3$ by causality. The expectations in perturbative QCD (orange hatched region), chiral perturbation theory (blue dashed curve) and the Stefan-Boltzmann limit (orange dotted line) are shown for comparison. }
    \label{fig:trace-anomaly}
\end{figure}

Two additional quantities that provide information about the nature of high-isospin-density matter are the polytropic index \cite{Annala:2019puf} and the trace anomaly \cite{Fujimoto:2022ohj} defined by
\begin{eqnarray}
    \gamma=\frac{\epsilon}{p}c_s^2 \label{eq:polytropic-index}, \\
    \Delta=\frac{1}{3}-\frac{p}{\epsilon} \label{eq:trace-anomaly},
\end{eqnarray}
respectively. The behavior of these two quantities is shown in Figs.~\ref{fig:polytropic-index} and \ref{fig:trace-anomaly} and compared to the expectations of a free gas, $\chi$PT and pQCD in each case. As for $c_s$, the behaviour of $\gamma$ and $\Delta$ is similar to that seen in Ref.~\cite{Brandt:2022hwy}, but the current work extends the range of chemical potential significantly which reveals additional interesting features. In Ref.~\cite{Annala:2019puf}, it is suggested that the point at which the polytropic index decreases below 1.75 is a sign of quark degrees of freedom at large baryon chemical potential, i.e., the BCS state. In the case of isospin chemical potential, $\gamma$ decreases to this value at $\mu_I\sim 1.5 m_\pi$, corresponding approximately to the position of the peak seen in the normalized energy density (Fig.~\ref{fig:energy-density-ratio}). 
The trace anomaly is clearly seen to be negative at intermediate $\mu_I$ in Fig.~\ref{fig:trace-anomaly}, as is suggested to be consistent with neutron star observations in Ref.~\cite{Fujimoto:2022ohj}. As for the quantities above, the results from the two lattice ensembles are in agreement for both the trace anomaly and the polytropic index.
A robust conclusion from the study of these transport quantities is that large isospin chemical potential is needed before the expected asymptotic behavior sets in. At least for the case of isospin chemical potential, the use of pQCD to describe the  behavior seen in the LQCD calculations requires $\mu_I \agt 10 m_\pi \sim 2$ GeV at a minimum.

\section{Summary and outlook}
\label{sec:summary}

In this work, a new, more efficient method of computing maximal-isospin, multi-pion correlation functions is presented. Using this method, we have calculated all $n$-$\pi^+$ correlation functions for $n \leq 6144$, extending such calculations of many-pion systems into regions of larger isospin chemical potential than have been previously achieved. Exploring such high-density and high-energy correlation functions presents its own suite of challenges owing to the range of numerical scales spanned by the correlation functions. Even on the same timeslice, correlation functions can vary by many orders of magnitude across configurations, leading to an effective breakdown of the applicability of the Central Limit Theorem. The analysis presented here overcomes this by making the empirically-driven assumption that the distributions of correlation functions across gauge configurations are log-normal, which allows the incorporation of more information about the LQCD data than just the sample mean and variance of the correlation functions.  With this assumption, it becomes possible to extract energies and chemical potentials from the LQCD correlation functions, which smoothly interpolate between theoretical predictions from chiral perturbation theory and perturbative QCD for  low- and high-isospin density systems, respectively. The speed of sound computed in this medium  exceeds the ideal gas limit over a large range of $\mu_I$, reaching a maximum of $c_s^2\sim 0.6$ at $\mu_I/m_\pi\sim 2$. This result is in agreement with the results of Ref.~\cite{Brandt:2022hwy} but extends over a larger range of chemical potential, lower temperatures, and to a  finer discretization scale. The isospin chemical potential is implemented through the  grand canonical partition function in Ref.~\cite{Brandt:2022hwy} and therefore the systematic uncertainties in that calculation are very different from those in this work, making the broad agreement seen more significant.  The speed of sound and other properties of the medium indicate that the asymptotic agreement with perturbative QCD expectations requires large values of the isospin chemical potential, ${\mu_I\gtrsim 2\text{ GeV}}$.

In this exploratory study, calculations have been performed at only a single set of quark masses and lattice spacing. The results show qualitative agreement with expectations, but understanding this system at a more precise level will require the use of additional ensembles with multiple lattice spacings, quark masses, and with other spatial and temporal extents in order to properly quantify the effects of these parameters on the calculation. Lattice cutoff effects are of particular concern since the maximum chemical potential reached in the calculations presented here comes close to the lattice cutoff scale used in this work.

Beyond systems of many pions, the methods developed here could also be used in applications to other systems of mesons, including systems of kaons and/or pions, and systems with non-zero momentum.
The concepts of symmetry and representation theory explored here to construct the algorithm for many-pion contractions can potentially be applied more broadly to baryonic systems. In addition, the success of log-normality in enabling analysis of many-pion systems points to the general observation that there is more information in the distributions of correlation functions than just their central values \cite{Guagnelli:1990jb,MJSPC,endres_noise_2011,DeGrand:2012ik,Beane:2009kya,Beane:2009gs,Beane:2009py,Wagman:2016bam,Wagman:2017gqi,Detmold:2018eqd,Davoudi:2020ngi,Yunus:2022wuc,Yunus:2023dka}, and using this information can allow the extraction of physical results even when the distributions of correlation functions are far from the regime of applicability of the Central Limit Theorem. 

\section*{Author Contribution Statement} 
 RA, WD, and FRL developed the algorithms and software for correlation function calculations, performed the numerical analysis, and prepared the manuscript; ZD, WD, MI, AP, RP, PES, and MLW  contributed to resource acquisition and propagator calculations and provided critical feedback on the manuscript.

\acknowledgements{
We are grateful to Massimo Mannarelli, Krishna Rajagopal, and Sanjay Reddy for discussions and to Balint Jo\'o for assistance with the generation of the  gauge configurations used in this work.
The calculations were performed using an allocation from the Innovative and Novel Computational Impact on Theory and Experiment (INCITE) program using the resources of the Oak Ridge Leadership Computing Facility located in the Oak Ridge National Laboratory, which is supported by the Office of Science of the Department of Energy under Contract DE-AC05-00OR22725.
This research also used resources of the National Energy Research Scientific Computing Center (NERSC), a U.S. Department of Energy Office of Science User Facility located at Lawrence Berkeley National Laboratory, operated under Contract No. DE-AC02-05CH11231.

This work is supported by the National Science Foundation under Cooperative Agreement PHY-2019786 (The NSF AI Institute for Artificial Intelligence and Fundamental Interactions, http://iaifi.org/) and by the U.S.~Department of Energy, Office of Science, Office of Nuclear Physics under grant Contract Number DE-SC0011090. 
RA and WD are also supported by the U.S.~Department of Energy SciDAC5 award DE-SC0023116. 
FRL acknowledges financial support from the Mauricio and Carlota Botton Fellowship. 
ZD is supported by the Maryland Center for Fundamental Physics and the College of Computer, Mathematical, and Natural Sciences at the University of Maryland, College Park.
MI is partially supported by the Quantum Science Center (QSC), a National Quantum Information Science Research Center of the U.S. Department of Energy.
PES is also supported by the U.S. DOE Early Career Award DE-SC0021006. 
AP and RP acknowledge support from Grant CEX2019-000918-M and the project PID2020-118758GB-I00, financed by the Spanish MCIN/ AEI/10.13039/501100011033/, and from the EU STRONG-2020 project under the program H2020-INFRAIA-2018-1 grant agreement no. 824093.
This manuscript has been authored by Fermi Research Alliance, LLC under Contract No. DE-AC02-07CH11359 with the U.S. Department of Energy, Office of Science, Office of High Energy Physics. 

This work made use of Chroma \cite{Edwards:2004sx}, QDPJIT \cite{Wint1405:Framework}, QUDA \cite{Clark:2010,Clark:2016rdz}, JAX~\cite{jax2018github}, NumPy~\cite{numpy}, SciPy~\cite{2020SciPy-NMeth}, and matplotlib~\cite{Hunter:2007}.
}

\appendix
\section{Numerical precision tests}\label{sec:numeric-checks}

In order to ensure that double precision floating-point numbers are sufficient for the calculation of the many-pion correlation functions, we performed several checks comparing quantities computed using double precision to the same quantities computed using higher-precision floating-point numbers. These checks are presented for the A ensemble, but similar conclusions can be drawn for the numerical stability of the calculations on the B ensemble.

As a first check, we compared our method against other methods using $12 \times 12$ pion blocks using propagators from a single point source. In particular, we implemented three pre-existing  algorithms for comparison: naive Wick contractions, direct computation of the traces using Eq.~\eqref{eq:correlator-sum-trace}, and the recursive method described in Ref.~\cite{Detmold:2010au}. We found agreement between all methods within numerical precision, provided that high-precision floating-point numbers were used in the other methods (our method gave indistinguishable results at double precision as at higher-precision).

\begin{figure}[t!]
    \centering
    \vspace{-0.1cm}
    \includegraphics[width=\linewidth]{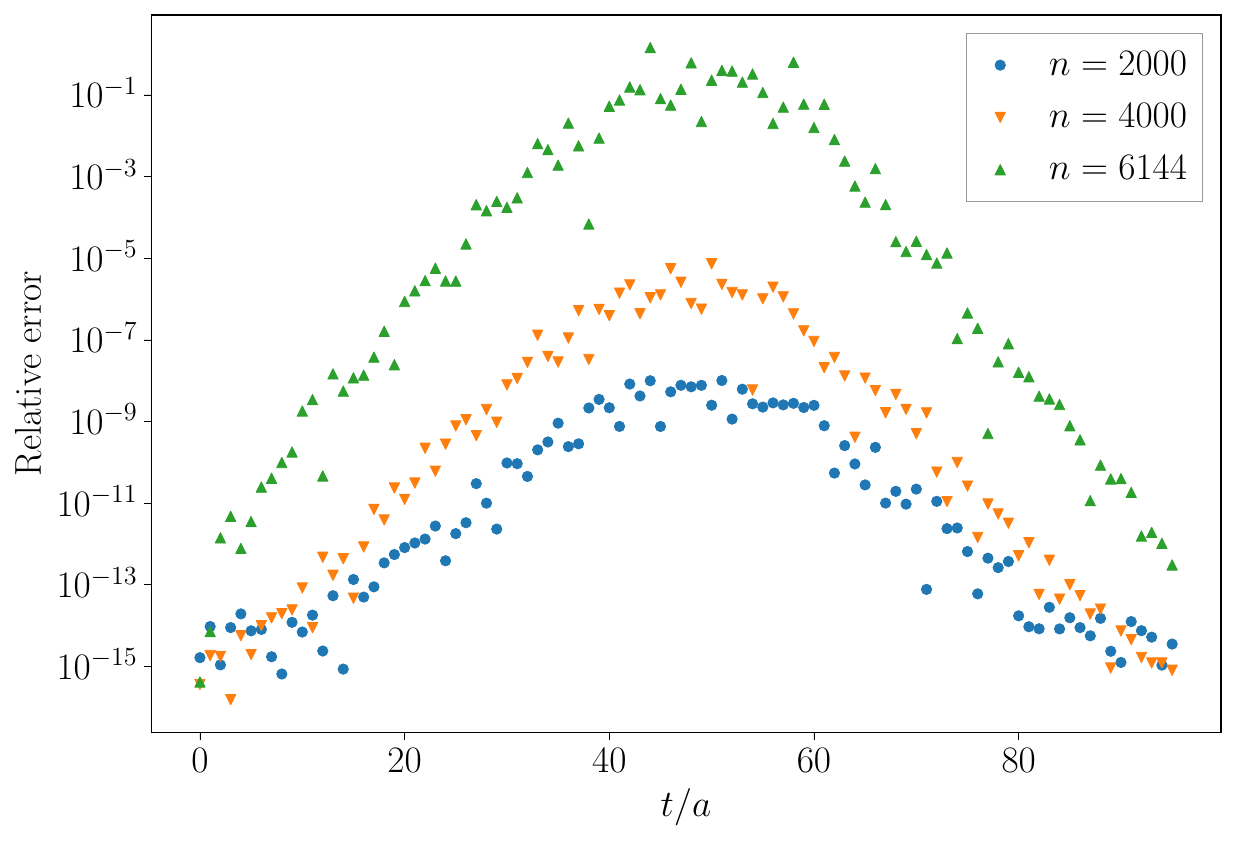}
    \caption{Relative errors on the eigenvalues of the pion block from finite precision in the SVD on the A ensemble. Errors are computed via $\epsilon = (x_i - x_i^{(\text{true})}) / x_i^{(\text{true})}$, where $x_i^{(\text{true})}$ is the double-double precision result.}
    \label{fig:eigenvalue-precision-error}
\end{figure}

\begin{figure}[t!]
    \centering
    \vspace{-0.1cm}
    \includegraphics[width=\linewidth]{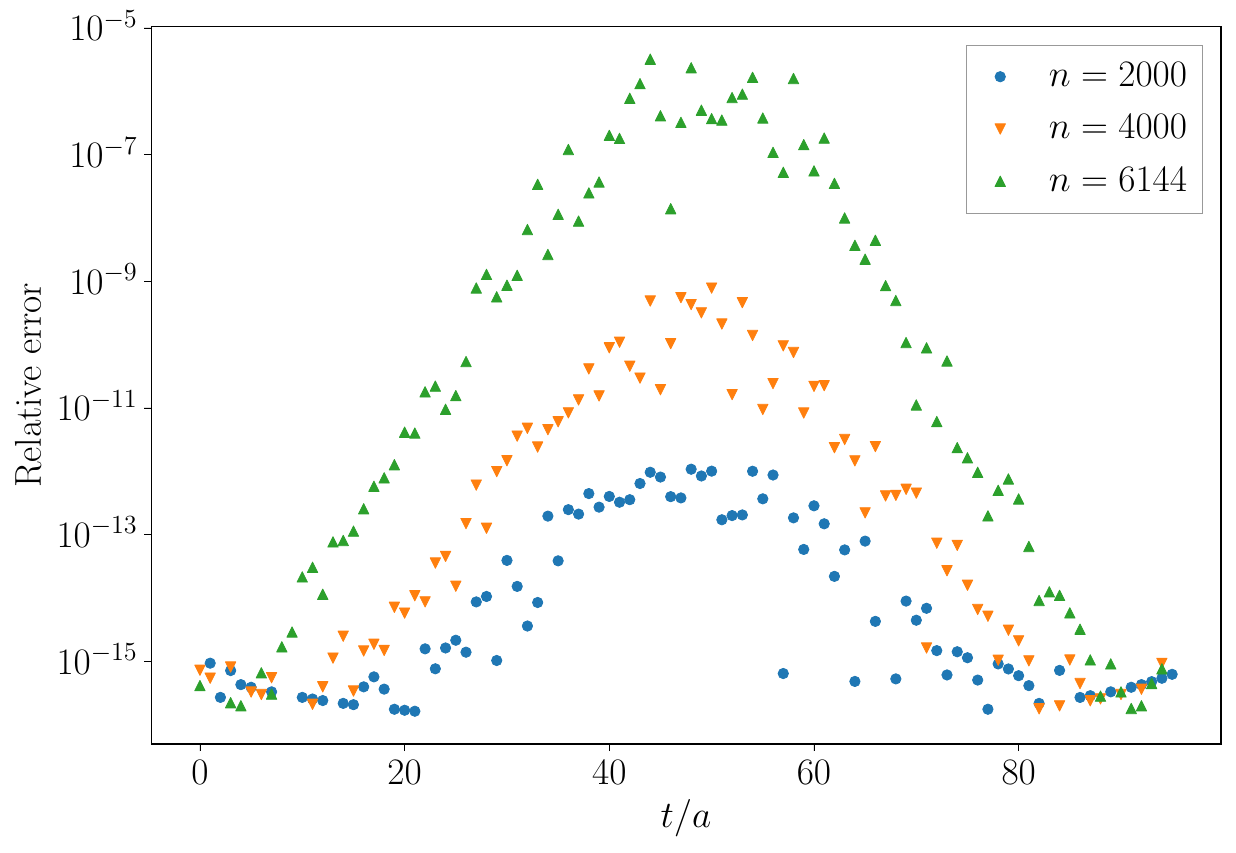}
    \caption{Precision errors on the logarithm of the correlation functions from finite precision in the SVD on the A ensemble. Details are as in Fig.~\protect\ref{fig:eigenvalue-precision-error}.
    }
    \label{fig:correlator-precision-errors}
\end{figure}

Next, we turned to the evaluation of correlation functions from the $6144 \times 6144$ pion block. The primary area of numerical concern is in the computation of the eigenvalues of the pion block, since round-off error in the SVD computation could reduce the accuracy of the smallest singular values, which could in turn render the correlation function computations inaccurate. In order to test the accuracy of the SVD at double precision, we performed single-configuration tests at higher precision. For the high-precision SVDs, we used {\tt GenericLinearAlgebra.jl} \cite{GenericLinearAlgebra} combined with {\tt MultiFloats.jl} \cite{MultiFloats} using 2 and 3 double-precision floats to emulate higher-precision floating-point numbers. We observed no difference in results at double precision between the 2- and 3-double precision SVDs, indicating that double-double precision is sufficient for the calculations in this work. The relative differences between the double-precision and double-double precision results for the eigenvalues, $x_n$, are shown in Fig.~\ref{fig:eigenvalue-precision-error}, while the relative differences for the correlation functions are shown in Fig.~\ref{fig:correlator-precision-errors}. Notably, the relative errors on some of the individual eigenvalues are large, reaching ${\cal O}(10\%)$, but these large errors do not propagate through to the correlation functions, which all have relative errors under 1 part in $10^5$.


\section{Cumulant expansion}\label{sec:cumulants}
The method of cumulants relies on the fact that for any random variable $X$ with finite moments, one may expand
\begin{equation}
    \label{eq:cumulant-expansion}
    \log \braket{e^X} = \sum_{n = 1}^\infty \frac{\kappa_n}{n!}\,,
\end{equation}
where $\kappa_n$ are the \emph{cumulants}, or connected correlation functions, of $X$, with the first few given by
\begin{align}
    \kappa_1 &= \mu = \braket{X}\,, \\
    \kappa_2 &= \braket{(X - \mu)^2}\,, \\
    \kappa_3 &= \braket{(X - \mu)^3} \,, \\
    \kappa_4 &= \braket{(X - \mu)^4} - 3 \braket{(X - \mu)^2}^2\,.
\end{align}
Applying this method to the problem at hand, the correlation function $C_n$ can be estimated by first estimating the first $N_\kappa$ cumulants $\kappa_1, \dots, \kappa_{N_\kappa}$ of $\log C_n$, and then combining these estimates using Eq.~\eqref{eq:cumulant-expansion} to obtain an estimate of $C_n$. For the case of $N_\kappa = 2$, this reduces exactly to the assumption of log-normality as discussed in Sec.~\ref{sec:log-normal}, but for higher $N_\kappa$, the use of cumulants provides a systematically-improvable method for estimating the correlation functions.

\begin{figure}[t!]
    \centering
    \includegraphics[width=\linewidth]{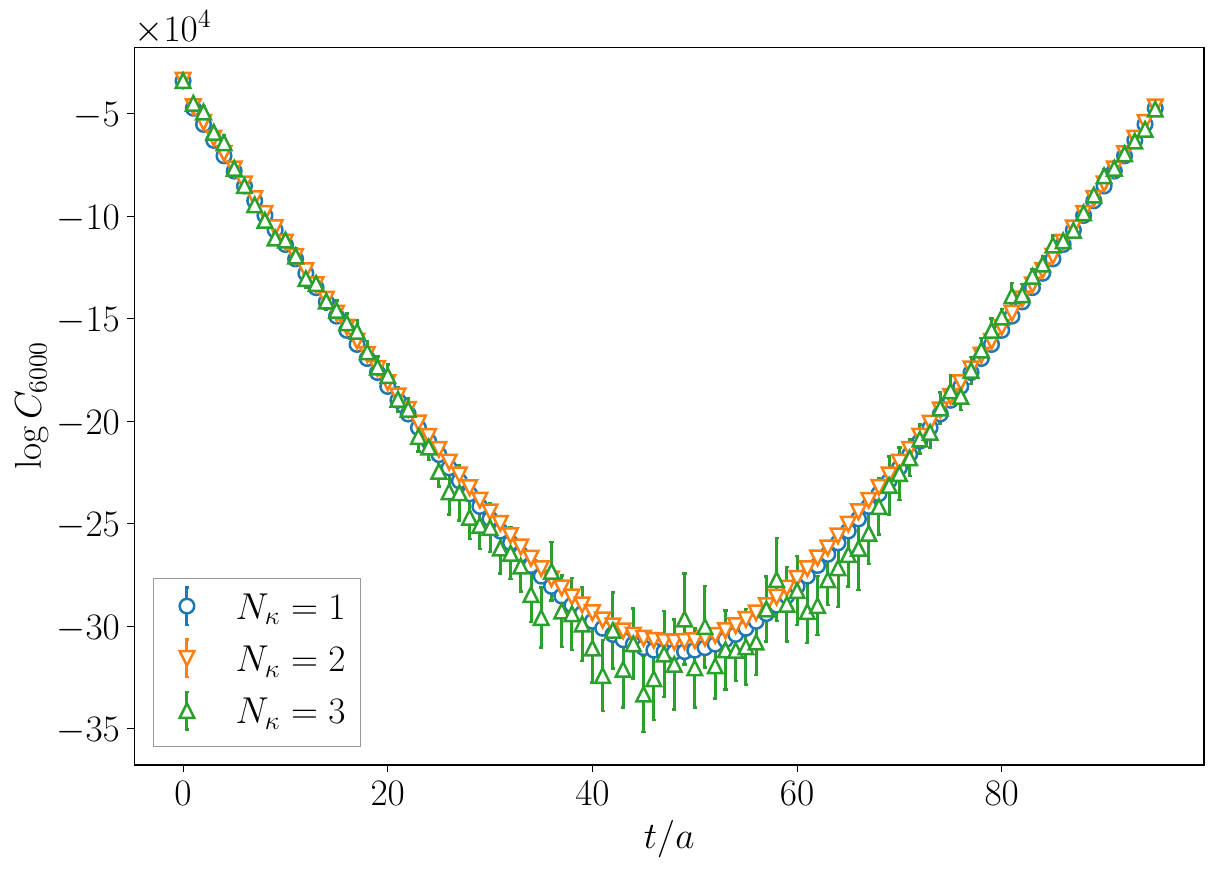}
    \caption{Cumulant corrected correlation function for ${n=6000}$ and for three different cumulant truncations on the A ensemble.}
    \label{fig:cumulant-correction}
\end{figure}

A qualitative picture of the effects of truncating the cumulant expansion can be seen in Fig.~\ref{fig:cumulant-correction}, where we show the value of the correlation function obtained after truncating at the first, second, and third order. The effect of the second-order cumulant (i.e., the variance) is small, but significant; meanwhile the third-order truncation is consistent with the second-order truncation but with significantly larger statistical uncertainties. The results are shown for the A ensemble, but similar behavior is seen on the B ensemble. As has been noted in Ref.~\cite{endres_noise_2011}, the cumulant expansion exhibits a bias-variance trade-off, with higher-order expansions being less biased but more noisy. In practice this means that higher-order cumulants do not improve the analysis at the current level of statistics, which is also consistent with the fact that we have been unable to detect statistical violations of log-normality.

\section{Details of data presentation}
\label{app:plotting}
The results shown in Figs.~\ref{fig:energy-density-ratio}, \ref{fig:speed-of-sound}, \ref{fig:polytropic-index}, and \ref{fig:trace-anomaly} arise from ${\cal O}(6000)$ densely packed points with uncertainties on both their $x$ and $y$ positions. This presents a challenge for accurately representing the data; this appendix contains details of the procedure used to generate these plots.
This procedure applies to any set of ordered data points $(x_i, y_i)$ along with associated uncertainties $(dx_i, dy_i)$. The algorithm is intended to create an envelope over the associated uncertainty ellipses defined by
\begin{align}
    x_i(\theta) &= x_i + dx_i \cos \theta, \\
    y_i(\theta) &= y_i + dy_i \sin \theta,
\end{align}
where $\theta \in [0, 2 \pi)$. Note that here the $x$ and $y$ uncertainties are treated as uncorrelated. The envelope of these ellipses is captured by sampling points along the ellipses and using linear interpolation to extend between the points. The exact procedure is most succinctly described via code, which is presented here in python using NumPy~\cite{numpy}:
\begin{lstlisting}[language=Python,keepspaces=false]
import numpy as np
def interpolate_fill_lines(x, y, *, xerr, yerr):
  thetas = np.linspace(0, 2 * np.pi, num=128)

  x_plt = np.concatenate(([x[0] - xerr[0]], x, \
                [x[-1] + xerr[-1]]), axis=-1)

  y_max = np.min(y - yerr) * np.ones_like(x_plt)
  y_min = np.max(y + yerr) * np.ones_like(x_plt)

  for theta in thetas:
    x1 = x + xerr * np.cos(theta)
    y1 = y + yerr * np.sin(theta)

    y_theta = np.interp(x_plot, x1, y1)
    y_max = np.maximum(y_max, y_theta)
    y_min = np.minimum(y_min, y_theta)

  return x_plt, y_min, y_max
\end{lstlisting}
The inputs \lstinline{x} and \lstinline{y} are both NumPy arrays containing the central values, while \lstinline{xerr} and \lstinline{yerr} indicate their respective uncertainties. The array \lstinline{x} is assumed to be monotonically increasing. The outputs of the function are a set of points $(x, y_\text{min}, y_\text{max})$, with $y_\text{min}$ indicating the lower boundary of the error band and $y_\text{max}$ indicating the upper boundary.

\begin{figure}[ht!]
    \centering
    \includegraphics[width=\linewidth]{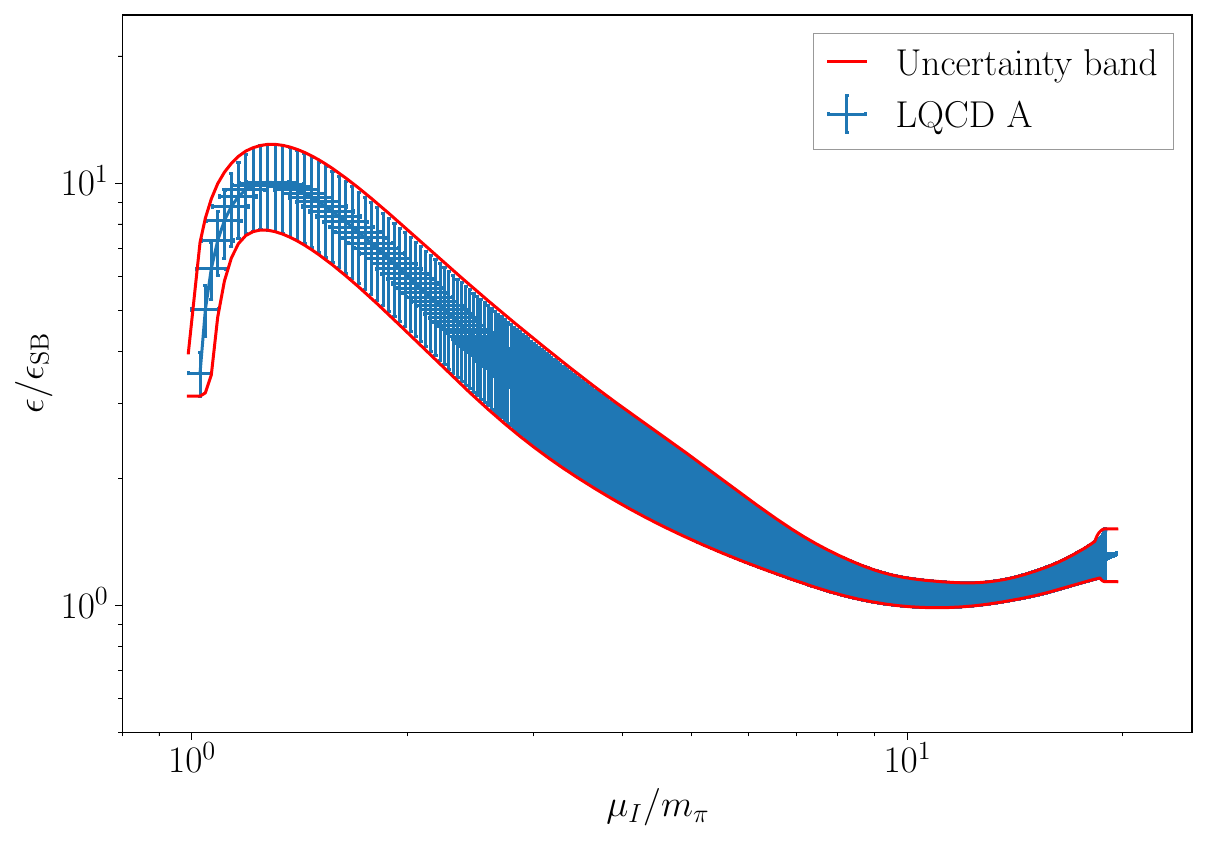}
    \caption{Comparison between the data points from the A ensemble used to generate Fig.~\ref{fig:energy-density-ratio} and the uncertainty bands generated as discussed in Appendix~\ref{app:plotting}. The horizontal and vertical extents of the blue crosses indicate the $x$ and $y$ uncertainties, respectively.}
    \label{fig:error-band}
\end{figure}

A qualitative picture of how the error bands generated from this procedure compared to the original data is shown in Fig.~\ref{fig:error-band} using the same data from ensemble A as in Fig.~\ref{fig:energy-density-ratio}. The blue crosses represent the original data, while the red lines indicate the upper and lower bounds of the uncertainty region displayed in Fig.~\ref{fig:energy-density-ratio} and can be seen to  tightly wrap the $x$ and $y$ uncertainties of the original data.

\bibliographystyle{utphys}
\bibliography{main}

\end{document}